\documentclass[paper]{JHEP3} 

\usepackage{epsfig,multicol}
\usepackage{amsmath}
\usepackage{amscd}
\usepackage{amsfonts}
\usepackage{amsthm}

\def\IN{\mathbb{N}}
\def\IZ{\mathbb{Z}}
\def\IR{\mathbb{R}}
\def\ID{\mathbb{D}}

\newcommand{\pp}{{=\!\!\!|}}
\newcommand\fverb{\setbox\pippobox=\hbox\bgroup\verb}
\newcommand\fverbdo{\egroup\medskip\noindent%
            \fbox{\unhbox\pippobox}\ }
\newcommand\fverbit{\egroup\item[\fbox{\unhbox\pippobox}]}
\newbox\pippobox

\title{The world-sheet description of A and B branes revisited}

\author{Alexander Sevrin, Wieland Staessens\thanks{Aspirant FWO} and Alexander Wijns\thanks{Address from October 1, 2007
on: Science Institute, University of Iceland,
Dunhaga 3, 107 Reykjav\'\i k, Iceland.}\\
Theoretische Natuurkunde, Vrije Universiteit Brussel and\\The International Solvay Institutes\\
Pleinlaan 2, B-1050 Brussels, Belgium \\
E-mail:  \email{Alexandre.Sevrin@vub.ac.be}, \email{awijns@tena4.vub.ac.be},
\email{Wieland.Staessens@vub.ac.be}}

\abstract{We give a manifest supersymmetric description of A and B
branes on K\"ahler manifolds using a completely local $N=2$ superspace formulation
of the world-sheet nonlinear $\sigma$-model in the presence of a
boundary. In particular, we show that an $N=2$ superspace
description of type A boundaries is possible, at least when the
background is K\"ahler. This leads to an elegant
and concrete setting for studying coisotropic A branes. Here, an
important role is played by the boundary potential, whose precise
physical meaning remains to be fully understood. Duality transformations relating A
and B branes in the presence of isometries are studied as well.}


\begin{document}
\setcounter{equation}{0}

%
%

\section{Introduction} \label{introduction}
Non-linear $ \sigma $-models in two dimensions with an $N=(2,2)$
supersymmetry, \cite{Alvarez-Gaume:1981hm}--\cite{Howe:1985pm}, play
a central role in the description of type II superstrings in the
absence of R-R fluxes. The interest in these models was recently
rekindled as well in the physics as in the mathematics community.
For physicists, these models allow for the study of
compactifications in the presence of non-trivial NS-NS fluxes, while
for mathematicians the models provide a concrete realization of
generalized complex geometries. A full off-shell supersymmetric
description clarifies the geometry behind these models. The case
without boundaries has been studied for more than two decennia and
has recently been fully solved in \cite{Lindstrom:2005zr} (building
on results in {\em e.g.}~\cite{Buscher:1987uw}--\cite{Bogaerts:1999jc}).
Formulating the model in $N=(2,2)$ superspace allows one to encode
the whole (local) geometry in a single scalar function, the Lagrange
density. The Lagrange density is a  function of scalar superfields
satisfying certain constraints. Only three types of superfields are
needed \cite{Lindstrom:2005zr}, \cite{MS}: chiral, twisted chiral
and semi-chiral superfields.

However, when dealing with D-branes one needs to confront $N=(2,2)$
non-linear $ \sigma $-models with boundaries. The presence of
boundaries breaks the $N=(2,2)$ supersymmetry down to an $N=2$
supersymmetry. While a lot of attention has been paid to these
models \cite{Ooguri:1996ck}--\cite{Lindstrom:2002jb}, their full
description in $N=2$ superspace has not been given yet.

In the present paper we open this study with the simplest case: A
and B branes on K\"ahler manifolds. While it is not too hard to
formulate B branes in $N=2$ superspace \cite{Koerber:2003ef}, type A
branes remained enigmatic up till now. As their boundary conditions
appear at first sight to be
incompatible with the complex structure associated with the
$N=(2,2)$ bulk supersymmetry, one expects a superspace
formulation to be subtle.

An important additional motivation for finding an $N=2$ world-sheet
superspace description of A branes is that it provides a new 
concrete setting for studying coisotropic branes. Indeed, in
\cite{Kapustin:2001ij} it was realized that in addition to the usual
type A branes wrapping lagrangian cycles, for consistency with
mirror symmetry which exchanges A and B branes, one should also
include so-called coisotropic branes. Their properties were already
established in \cite{Kapustin:2001ij} and later re-derived from a
world-sheet point of view in \cite{Lindstrom:2002jb}. So far, the
only concrete examples appearing in the literature are maximally
coisotropic branes on $T^4$ \cite{Kapustin:2001ij,Aldi:2005hz} and
$K3$ \cite{Kapustin:2006pk}, and coisotropic branes wrapping
5-cycles on $T^6$, $T^6/\IZ_2 \times \IZ_2$ and $T^2 \times K3$
\cite{Font:2006na}.

In the next section we revisit the $N=(1,1)$ supersymmetric
non-linear $ \sigma $-models in the presence of boundaries thereby
clarifying some remaining problems. In section 3 we construct $N=2$
superspace. We show that changing from A to B boundary conditions
amounts to interchanging chiral superfields and twisted chiral ones
and vice versa. In section 4 we give a detailed description of type
A branes followed by a similar description of type B branes in
section 5. When certain isometries are present, chiral superfields
can be dualized into twisted chiral superfields and vice versa. In
section 6 we study these duality transformations in the presence of
boundaries. We end with conclusions and an outlook. The study of
general non-linear $ \sigma $-models with boundaries -- involving chiral, twisted
chiral and semi-chiral fields simultanously -- will appear elsewhere
\cite{wip}.

\section{From $N=(1,1)$ to $N=1$}
A non-linear $ \sigma $-model (with $N\leq(1,1)$) on some target manifold $ {\cal M}$ is
characterized by a metric
$g_{ab}$ and a closed 3-form $T_{abc}$ (known as the torsion, the Kalb-Ramond 3-form or the NS-NS
flux) on $ {\cal M}$. The action in $N=(1,1)$ superspace is simply\footnote{Our conventions are
given in appendix A.},
\begin{eqnarray}
{\cal S}=2\int d^2 \sigma \, d^2 \theta \,D_+X^aD_-X^b\left(g_{ab}+b_{ab}\right),\label{an11}
\end{eqnarray}
where we used the locally defined 2-form potential $b_{ab}$ for the torsion,
\begin{eqnarray}
T_{abc}=- \frac{3}{2}\, \partial _{[a}b_{bc]}.
\end{eqnarray}

We introduce a boundary\footnote{As far as we know, the first place where superspaces with 
boundaries were introduced and used was in \cite{Itoyama:1987qz}.} at 
$ \sigma =0$ ( $ \sigma \geq 0$ ) and $ \theta ^+= \theta ^-$. This
breaks the invariance under translations in both the $ \sigma $ and the
$  \theta' \equiv \theta ^+- \theta ^-$ direction. Put differently, the presence of a boundary
breaks the $N=(1,1)$ supersymmetry to an $N=1$ supersymmetry. We introduce the derivatives,
\begin{eqnarray}
D\equiv D_++D_-,\qquad D'\equiv D_+-D_-,
\end{eqnarray}
which satisfy,
\begin{eqnarray}
D^2=D'{}^2=- \frac{i}{2} \partial _ \tau ,\qquad \{D,D'\}=-i\, \partial _ \sigma .
\end{eqnarray}
Using this, one verifies that
\begin{eqnarray}
-D\,D'=2 D_+\,D_- + \frac{i}{2}\,\partial _ \sigma,\label{ddsig}
\end{eqnarray}
which allows us to write a manifest $N=1$ supersymmetric lagrangian,
\begin{eqnarray}
{\cal S}=-\int d^2 \sigma \, d \theta
\,D'\left(D_+X^aD_-X^b\left(g_{ab}+b_{ab}\right)\right),\label{an1}
\end{eqnarray}
which -- because of eq.~(\ref{ddsig}) -- differs from the action in the absence of boundaries,
eq.~(\ref{an11}), by a boundary term \cite{Lindstrom:2002mc}, \cite{Koerber:2003ef}. Working out the
$D'$ derivative yields the action in $N=1$ boundary superspace obtained in \cite{Koerber:2003ef},
\begin{eqnarray}
{\cal S}&=&\int d^2 \sigma \, d \theta\Big(
i\,g_{ab}\,DX^a \partial _ \tau  X^b-2i\,g_{ab}\, \partial _ \sigma X^a D'X ^b+\, 2i\,b_{ab}\,
\partial _ \sigma X^a DX^b \nonumber\\
&&- 2 \,g_{ab}\, D'X ^a \nabla D'X ^b+2\,T_{abc} \,D'X ^aDX^bDX^c- \frac{2}{3} \,T_{abc}\,
D'X ^a D'X ^b D'X ^c\Big),\label{an12}
\end{eqnarray}
where,
\begin{eqnarray}
\nabla  D'X ^a\equiv D D'X ^a+ \left\{ {}^{\, a}_{bc} \right\}DX^b D'X ^c,
\end{eqnarray}
and both $X^a$ and $D'X^a$ should now be viewed as independent $N=1$ superfields.
Note that when $b_{ab}= \partial _a A_b- \partial _b A_a$, we can rewrite eq.~(\ref{an12}) as,
\begin{eqnarray}
{\cal S}&=&\int d^2 \sigma \, d \theta\Big(
i\,g_{ab}\,DX^a \partial _ \tau  X^b-\,2i\,g_{ab}\, \partial _ \sigma X^a D'X^b - 2 \,g_{ab}\,
D'X ^a \nabla D'X ^b\Big) \nonumber\\
&&+\, 2i\,\int d \tau d \theta \, A_a\,DX^a.
\label{an122}
\end{eqnarray}

Varying the action eq.~(\ref{an1})\footnote{Where one uses that
$\int d^2 \sigma d \theta D'\,D_\pm=-(i/2)\,\int d \tau d \theta$. } or
eq.~(\ref{an12}) yields a boundary term,
\begin{eqnarray}
\delta {\cal S}\big| _{boundary}= -2i\,\int d \tau d \theta \,\delta X^a\left(
g_{ab} \,D'X ^b-\, b_{ab}\,DX^b\right).\label{var1}
\end{eqnarray}
This boundary term will only vanish if suitable boundary conditions are imposed. In order to do so
we introduce a (1,1) tensor $R(X)^a{}_b$ \cite{Ooguri:1996ck}, \cite{Albertsson:2001dv},
\cite{Albertsson:2002qc},
\cite{Koerber:2003ef} which satisfies,
\begin{eqnarray}
R^a{}_c\,R^c{}_b= \delta ^a_b,
\end{eqnarray}
and projection operators $ {\cal P}_\pm$,
\begin{eqnarray}
{\cal P}_\pm^a{}_b\equiv \frac{1}{2}\left( \delta ^a_b\pm R^a{}_b\right).
\end{eqnarray}
With this we impose Dirichlet boundary conditions,
\begin{eqnarray}
{\cal P}_-^a{}_b\, \delta X^b=0.\label{dbc1}
\end{eqnarray}
Using eq.~(\ref{dbc1}), one verifies that the boundary term eq.~(\ref{var1}) vanishes, provided one
imposes Neumann boundary conditions,
\begin{eqnarray}
{\cal P}_{+ba}\,D'X^b= {\cal P}_+^b{}_a\,b_{bc}\,DX^c,\label{nbc1}
\end{eqnarray}
as well. If in addition we assume -- for which at this point, as we will demonstrate in an
example later on, there is no necessary reason -- that, 
\begin{eqnarray}
g_{ac}R^c{}_b=g_{bc}R^c{}_a,
\end{eqnarray}
or $R_{ab}=R_{ba}$, then we can rewrite eq.~(\ref{nbc1}) as,
\begin{eqnarray}
{\cal P}^a_+{}_b\,D'X^b= {\cal P}_+^a{}_c\,b^c{}_{d} {\cal P}_+^d{}_b\,DX^b.\label{xxnbc1}
\end{eqnarray}
Invariance of the
Dirichlet boundary conditions under what remains of the super-Poincar\'e transformations implies
that on the boundary,
\begin{eqnarray}
{\cal P}_-^a{}_b\,DX^b={\cal P}_-^a{}_b\, \partial _ \tau X^b=0,\label{ghghg}
\end{eqnarray}
hold as well. Using $D^2=-i/2\, \partial _ \tau $, we get from eq.~(\ref{ghghg}) the integrability
conditions\footnote{
Out of two $(1,1)$ tensors $R^a{}_b$ and $
S^a{}_b$, one constructs a $(1,2)$ tensor
${\cal N}[R,S]^a{}_{bc}$, the Nijenhuis tensor, as ${\cal N}[R,S]^a{}_{bc}=
R^a{}_dS^d{}_{[b,c]}+R^d{}_{[b}S^a{}_{c],d}+R\leftrightarrow S$.},
\begin{eqnarray}
0={\cal P}_+^d{}_{[b}{\cal P}_+^e{}_{c]} {\cal P}^a_+{}_{d,e}=
-\, \frac{1}{2}\, {\cal P}_-^a{}_e \,{\cal N}^e{}_{bc}[R,R].\label{ic11}
\end{eqnarray}
These conditions guarantee the existence of adapted coordinates $X^{\hat a}$, $\hat a
\in\{p+1,\cdots ,d\}$, with $p\leq d$ the rank of $ {\cal P_+}$ such that the Dirichlet boundary
conditions, eq.~(\ref{dbc1}) are simply given by,
\begin{eqnarray}
X^{\hat a}=\mbox{ constant}, \qquad \forall \,\hat a\in\{p+1,\cdots, d\}.\label{dbc9}
\end{eqnarray}
Writing the remainder of the coordinates as $X^{\check a}$, $\check a \in\{1,\cdots, p\}$, we get
the Neumann boundary conditions, eq.~(\ref{nbc1}), in our adapted coordinates,
\begin{eqnarray}
g_{\check a b}\,D'X^b=b_{\check a\check b}\,DX^{\check b},\label{nbc9}
\end{eqnarray}
where $b$ is summed from 1 to $d$ and where we used that $DX^{\hat b}$ vanishes on the
boundary. Concluding, the action eq.~(\ref{an1}) together with the boundary
conditions eqs.~(\ref{dbc9}) and (\ref{nbc9}), describe open strings in the presence of a
D$p$-brane whose position is determined by eq.~(\ref{dbc9}).

Let us end this section with an example. We start with a very simple
configuration consisting of a D2-brane on a 2-torus with coordinates $X^1$ and $X^2$ in the presence
of a $U(1)$ magnetic background\footnote{Whenever $b_{ab}$ is closed, we will denote it by
$F_{ab}$.} $F_{12}=F(X^1)$ (note that $ \partial _2F=0$). The action is,
\begin{eqnarray}
{\cal S}_{D2}&=&-\int\,d^2 \sigma d \theta D'\,\Big(
D_+X^1D_-X^1+D_+X^2D_-X^2+ \nonumber\\
&& F(X^1)\,\big(D_+X^1D_-X^2-D_+X^2D_-X^1\big)
\Big),
\end{eqnarray}
and the boundary conditions are Neumann in all directions,
\begin{eqnarray}
D'X^1=+\,F(X^1)\, DX^2,\qquad D'X^2=-\,F(X^1)\, DX^1.
\end{eqnarray}
Making a T-duality transformation along the $X^2$ direction\footnote{A simple way to do this
is by gauging the isometry $X^2\rightarrow X^2+\ \mbox{constant}$ and -- using Lagrange 
multipliers -- imposing that the gauge fields are pure gauge. Integrating over the gauge fields 
yields the T-dual model, see {\em e.g.}~\cite{Alvarez:1996up}.} yields a D1-brane with action,
\begin{eqnarray}
{\cal S}_{D1}&=&-\int\,d^2 \sigma d \theta D'\Big(
(1+F^2)\,D_+X^1D_-X^1+D_+X^2D_-X^2 \nonumber\\
&&+F\,\big(D_+X^1D_-X^2+D_+X^2D_-X^1\big)
\Big),\label{zyz}
\end{eqnarray}
and boundary conditions,
\begin{eqnarray}
&&\delta X^2=0, \nonumber\\
&& (1+F^2)D'X^1+F\,D'X^2=0.\label{qqqs}
\end{eqnarray}
Comparing eq.~(\ref{zyz}) to eq.~(\ref{an1}), we read off the (flat) metric: $g_{11}=1+F^2$,
$g_{12}=F$ and
$g_{22}=1$. Comparing the boundary conditions eq.~(\ref{qqqs}) with eqs.~(\ref{dbc1}) and
(\ref{nbc1}), we get $R^1{}_1=1$, $R^1{}_2=2F/(1+F^2)$, $R^2{}_1=0$ and $R^2{}_2=-1$. One verifies 
that for this choice of $R^a{}_b$, $R_{ab}=R_{ba}$ holds. Note that we might as well 
have chosen $R^1{}_1=-R^2{}_2=1$ and $R^1{}_2=R^2{}_1=0$ which also reproduce the 
boundary conditions eq.~(\ref{qqqs}). However for this choice we have $R_{ab}\neq R_{ba}$.

The D1-brane configuration described here is
fairly standard. Indeed, take the Dirichlet boundary condition to be $X^2=0$ and change coordinates,
\begin{eqnarray}
Y^1=X^1,\qquad Y^2=X^2+A(X^1),\label{newcds}
\end{eqnarray}
where the potential $A(X^1)$ is defined by
$F(X^1)= \partial _1 A(X^1)$. In these coordinates the metric
becomes the standard one, $g_{ab}= \delta _{ab}$, and the D1-brane is defined by $Y^2=A(Y^1)$, 
where $Y^1$ assumes the role of worldvolume coordinate.
Taking a constant magnetic field, $F=\tan \theta $, we recognize the system as a straight D1-brane
rotated in the $Y^1Y^2$-plane over an angle $ \theta $ with respect to the $Y^1$-axis.

\section{N=2 superspace}
\subsection{$N=(2,2)$ supersymmetry in the absence of boundaries}
Even without boundaries, promoting the $N=(1,1)$ supersymmetry of the action in eq.~(\ref{an11}) to
an $N=(2,2)$ is a non-trivial operation which introduces a lot of additional geometric structure in
the model. The most general extra supersymmetry transformations -- consistent with
dimensions and Poincar\'e symmetry -- are of the form,
\begin{eqnarray}
\delta X^a= \varepsilon ^+\,J_+^a{}_b(X)\,D_+X^b+\varepsilon ^-\,J_-^a{}_b(X)\,D_-X^b,\label{tr22}
\end{eqnarray}
which requires the introduction of two (1,1) tensors $J_+$ and $J_-$. Requiring the supersymmetry
algebra to close {\em on-shell}, one finds that both $J_+$ and $J_-$ must be complex structures,
\begin{eqnarray}
&&J_\pm^a{}_c\,J_\pm^c{}_b=- \delta ^a_b, \nonumber\\
&& N[J_\pm,J_\pm]^a{}_{bc}=0.
\end{eqnarray}
Apart from requiring that the $N=(2,2)$ supersymmetry algebra is
satisfied, we have to demand that the action eq.~(\ref{an11}) is
invariant under the transformations eq.~(\ref{tr22}). This yields
additional conditions. The metric has to be hermitian with respect
to {\em both} complex structures\footnote{This implies the existence
of two two-forms $ \omega^\pm _{ab}=-\omega^\pm _{ba}=
g_{ac}J^c_\pm{}_b$. In general they are not closed. Using
eq.~(\ref{covconst}), one shows that $ \omega ^\pm_{[ab,c]}=\mp 2
J_\pm^d{}_{[a}T_{bc]d}=\mp (2/3)J^d_\pm{}_a
J^e_\pm{}_bJ^f_\pm{}_cT_{def}$, where for the last step we used the
fact that the Nijenhuis tensors vanish.},
\begin{eqnarray}
J_\pm^c{}_a\,J_\pm^d{}_b\,g_{cd}=g_{ab}\,.\label{hermit}
\end{eqnarray}
Furthermore, both complex structures have to be covariantly constant,
\begin{eqnarray}
0=\nabla_c^\pm \,J_\pm^a{}_b\equiv
\partial _c\,J_\pm^a{}_b+\Gamma^a_{\pm dc}J_\pm^d{}_{b}-
\Gamma^d_{\pm bc}J_\pm^a{}_d\,,\label{covconst}
\end{eqnarray}
with the connections $\Gamma_\pm$ given by,
\begin{eqnarray}
\Gamma^a_{\pm bc}\equiv  \left\{ {}^{\, a}_{bc} \right\} \pm
T^a{}_{bc}\,.\label{cons}
\end{eqnarray}
A complex manifold with the above additional properties is called
bihermitian. When the torsion vanishes, this type of geometry
reduces to the usual K\"ahler geometry.

When calculating the algebra explicitly one finds that the terms in
the algebra which do not close off-shell are proportional to the
commutator of the complex structures ${[}J_+,J_-{]}$. 
In order to obtain an off-shell closing formulation of the model, one
expects that $\ker {[}J_+,J_-{]}$ can be described without any
additional auxiliary fields while the description of
$\mbox{coker}{[}J_+,J_-{]}$ will require the introduction of new
auxiliary fields. This picture was already suggested in
\cite{Sevrin:1996jr} and \cite {Bogaerts:1999jc} (see also
\cite{Ivanov:1994ec}) and was shown in \cite{Lindstrom:2005zr} to be
correct. Roughly speaking one gets that when writing
$\ker{[}J_+,J_-{]}=\ker(J_+-J_-)\oplus\ker(J_++J_-)$,
$\ker(J_+-J_-)$ and $\ker(J_++J_-)$ resp.~can be integrated to
chiral and twisted chiral multiplets resp.~\cite{Gates:1984nk}.
Semi-chiral multiplets \cite{Buscher:1987uw} are required for the
description of $\mbox{coker}{[}J_+,J_-{]}$.

In the present paper we will focus on chiral and twisted chiral
multiplets, {\em i.e.}~we assume that $J_+$ and $J_-$
commute\footnote{As already mentioned in the introduction we
relegate the study of the most general case -- which includes the
semi-chiral superfields -- to a forthcoming paper \cite{wip}.}.
These fields in $N=(2,2)$ superspace (once more we refer to the
appendix for conventions) satisfy the constraints $\hat D_\pm
X^a=J^a_\pm{}_b\,D_\pm X^b$ where $J_+$ and $J_-$ can be
simultaneously diagonalized. When the eigenvalues of $J_+$ and $J_-$
have the same (the opposite) sign we have chiral (twisted chiral)
superfields. Explicitly, we get that chiral superfields $ X^ \alpha
$, $ \alpha \in \{1,\cdots, m\}$, satisfy,
\begin{eqnarray}
\hat D_\pm X^ \alpha =+i\,D_\pm X^ \alpha ,\qquad
\hat D_\pm X^{ \bar \alpha }=-i\,D_\pm X^{ \bar \alpha }.\label{defsufi1}
\end{eqnarray}
Twisted chiral superfields $X^ \mu $, $ \mu \in \{1,\cdots, n\}$ satisfy,
\begin{eqnarray}
\hat D_\pm X^ \mu =\pm i\, D_\pm X^ \mu ,\qquad
\hat D_\pm X^{ \bar \mu }=\mp i\,D_\pm X^{ \bar \mu }.\label{defsufi2}
\end{eqnarray}
The most general action involving these superfields is given by,
\begin{eqnarray}
{\cal S}=\int\,d^2 \sigma \,d^2\theta \,d^2 \hat \theta \, V(X, \bar X),
\end{eqnarray}
where the Lagrange density $V(X, \bar X)$ is an arbitrary real function of the chiral and twisted
chiral superfields. Passing to $N=(1,1)$ superspace and comparing the result to eq.~(\ref{an11}),
allows one to identify the metric and the torsion potential\footnote{Indices from the beginning of
the Greek alphabet, $ \alpha $, $ \beta $, $ \gamma $, ... denote chiral fields while indices from
the middle of the alphabet, $ \mu $, $ \nu $, $ \rho $, ... denote twisted chiral fields.},
\begin{eqnarray}
&&g_{ \alpha \bar \beta }=+V_{ \alpha \bar \beta },\qquad g_{ \mu \bar \nu }= -
V_{ \mu \bar \nu },\nonumber\\
&&b_{ \alpha \bar \nu }=-V_{ \alpha \bar \nu }, \qquad b_{ \mu \bar \beta  }=+V_{ \mu \bar \beta  },
\label{KRform}
\end{eqnarray}
where all other components of $g$ and $b$ vanish. When writing $V_{
\alpha \bar \beta }$, we mean $ \partial _ \alpha \partial_ { \bar
\beta }V$ etc. Note that when only one type of superfield is
present, the target manifold is K\"ahler, which is the case in which
we are presently interested. The case where both of them are
simultaneously present will be discussed elsewhere \cite{wip}.

\subsection{From $N=(2,2)$ to $N=2$}
We now assume that in the bulk -- far away from the boundary -- the model exhibits an $N=(2,2)$
supersymmetry as described in the previous subsection. We expect the boundary to break half of
the supersymmetries, so we will go from $N=(2,2)$ to $N=2$. In order to handle this we rewrite
eq.~(\ref{tr22}) as,
\begin{eqnarray}
\delta X^a= \varepsilon \,J^{(+)}{}^a{}_b\,DX^b+\varepsilon\, J^{(-)}{}^a{}_b\,D'X^b+
\varepsilon '\, J^{(-)}{}^a{}_b\,DX^b+\varepsilon ' \,J^{(+)}{}^a{}_b\,D'X^b,
\end{eqnarray}
where,
\begin{eqnarray}
&&\varepsilon \equiv \frac{1}{2}( \varepsilon ^++ \varepsilon ^-),\qquad
\varepsilon ' \equiv \frac{1}{2}( \varepsilon ^+- \varepsilon ^-), \nonumber\\
&&J^{(\pm)}\equiv \frac{1}{2}\left( J_+\pm J_- \right).
\end{eqnarray}
Whenever the $ \varepsilon $ supersymmetry is preserved, one talks
about B-type boundary conditions, while preservation of the $
\varepsilon '$ supersymmetry corresponds to what are called A-type
boundary conditions. One sees that switching from B-type to A-type
amounts to replacing $ \varepsilon $ by $ \varepsilon '$ and $
J^{(\pm)}$ by $ J^{(\mp)}$. In $N=(2,2)$ superspace, B-boundary
conditions correspond to a boundary $ \theta '\equiv (\theta ^+-
\theta ^-)/2=0$ and $ \hat \theta '\equiv(\hat \theta ^+-\hat \theta
^-)/2=0$. A-type boundary conditions on the other hand correspond to
$ \theta '\equiv (\theta ^+- \theta ^-)/2=0$ and $ \hat \theta
'\equiv(\hat \theta ^++\hat \theta ^-)/2=0$. For B-type boundaries
we define,
\begin{eqnarray}
&&D\equiv D_++D_-,\qquad \hat D\equiv \hat D_++ \hat D_-, \nonumber\\
&& D'\equiv D_+-D_-,\qquad \hat D'\equiv \hat D_+- \hat D_-,\label{Aders}
\end{eqnarray}
where unaccented derivatives refer to translations in the invariant directions.
When dealing with A-type boundaries, the role of $\hat D$ and $\hat D'$ are interchanged.
For the moment we will
focus on B-type boundaries. Later on we will see that this does not present any restriction as
switching from one type of boundary conditions to another will just amount to interchanging
chiral for twisted chiral superfields and vice-versa.
The derivatives defined in eq.~(\ref{Aders}) satisfy,
\begin{eqnarray}
&&D^2=\hat D^2=D'{}^2=\hat D'{}^2=- \frac{i}{2} \partial _ \tau , \nonumber\\
&&\{D,D'\}=\{\hat D, \hat D'\}=-i \partial _ \sigma ,
\end{eqnarray}
and all other anti-commutators vanish.

Let us now turn to the superfields. In the bulk we had chiral, twisted chiral and semi-chiral
superfields. In the present paper we focus on chiral and twisted chiral superfields. From
eqs.~(\ref{defsufi1}) and (\ref{Aders}) we get for the chiral fields,
\begin{eqnarray}
&&\hat D X^ \alpha =+i DX^ \alpha ,\quad
\hat D X^ {\bar \alpha} =-i DX^{\bar \alpha}  \nonumber\\
&&\hat D' X^ \alpha =+i D'X^ \alpha ,\quad
\hat D' X^ {\bar \alpha} =-i D'X^{\bar \alpha} ,\label{bcf}
\end{eqnarray}
where $ \alpha,\ \bar \alpha  \in\{1,\cdots, m\}$.
Passing from $N=(2,2)$ -- parametrized by the Grassmann coordinates $ \theta $, $ \hat \theta $, 
$ \theta '$ and $ \hat \theta '$ -- to $N=2$ superspace --
parametrized by $ \theta $ and $ \hat \theta $ -- we get $X^ \alpha $, $X^{ \bar \alpha }$,
$D'X^ \alpha $ and $D'X^{ \bar \alpha }$ as $N=2$ superfields and they satisfy the constraints,
\begin{eqnarray}
&&\hat D X^ \alpha =+i\,D X^ \alpha ,\qquad
\hat D X^{ \bar \alpha }=-i\, D X^{ \bar \alpha }, \nonumber\\
&&\hat D \,D'X^ \alpha =+i\,D \,D'X^ \alpha- \partial _ \sigma X^ \alpha  ,\qquad
\hat D\,D' X^{ \bar \alpha }=-i\, D\,D' X^{ \bar \alpha }+ \partial _ \sigma X^{ \bar \alpha }.
\label{n2csf}
\end{eqnarray}

For twisted chiral superfields we get instead, when combining
eqs.~(\ref{defsufi2}) and (\ref{Aders}),
\begin{eqnarray}
&&\hat D X^ \mu =+i D'X^ \mu ,\quad
\hat D X^ {\bar \mu} =-i D'X^{\bar \mu},  \nonumber\\
&&\hat D' X^ \mu =+i DX^ \mu ,\quad
\hat D' X^ {\bar \mu} =-i DX^{\bar \mu} ,\label{btcf}
\end{eqnarray}
with $ \mu ,\ \bar \mu \in\{1,\cdots, n\}$. Passing again
from $N=(2,2)$ to $N=2$ superspace, we now get $X^ \mu  $, $X^{ \bar \mu  }$,
$D'X^ \mu  $ and $D'X^{ \bar \mu  }$ as $N=2$ superfields satisfying the constraints,
\begin{eqnarray}
&&\hat D X^{ \mu }=+i\, D'X^{ \mu },\quad
\hat D X^{ \bar  \mu }=-i\, D'X^{ \bar  \mu }, \nonumber\\
&& \hat D \,D'X^{ \mu }=- \frac{1}{2}\dot X^{ \mu },\quad
\hat D\,D' X^{ \bar  \mu }=+ \frac{1}{2}\dot X^{ \bar  \mu }.\label{n2tcsf}
\end{eqnarray}
So in $N=2$ superspace, the twisted chiral superfields $X^ \mu $ and $X^{ \bar \mu }$
are unconstrained superfields.

It is important to note that, had we used A-type boundaries instead
of B-type, we would have gotten exactly the same expressions but
with the roles of chiral and twisted chiral fields interchanged. We
will return to duality transformations interchanging chiral for
twisted chiral fields and vice versa in section \ref{sec:dual}.

Once more one immediately verifies that the difference between the fermionic measure
$D_+D_- \hat D_+\hat D_-$ and $D \hat D D' \hat D'$ is just a boundary term. So the (al)most
general $N=2$ invariant action which reduces to the usual action far away from the boundary
we can write down is,
\begin{eqnarray}
{\cal S}=\int d^2 \sigma\, d \theta d \hat \theta\, D' \hat D'\, V(X, \bar X),
\end{eqnarray}
where $V( X, \bar X)$ is an arbitrary function of the (bulk) superfields. In fact, when boundaries
are present, we can still generalize the previous by adding a boundary term,
\begin{eqnarray}
{\cal S}=\int d^2 \sigma\, d \theta d \hat \theta\, D' \hat D'\, V(X, \bar X)+
i\,\int d \tau \,d \theta d \hat \theta \,W( X, \bar X),
\end{eqnarray}
with $W( X , \bar X)$ an arbitrary function of the (bulk) superfields.

\section{Type A branes}
Type A branes on K\"ahler manifolds are described in terms of twisted chiral fields,
eqs.~(\ref{btcf}) and (\ref{n2tcsf}). The most general $N=2$ supersymmetric action we can write
down is,
\begin{eqnarray}
{\cal S}=\int d^2 \sigma d^2 \theta D'\hat D' \,V( X, \bar X)+ i\,\int d \tau d^2 \theta\,
W(X, \bar X).\label{aaa0}
\end{eqnarray}
with $V( X, \bar X)$ and $W( X, \bar X)$ arbitrary functions of the twisted chiral fields.
Working out the $ \hat D'$ and $D'$ derivatives using the constraints eq.~(\ref{btcf}) gives
the action in $N=2$ boundary superspace,
\begin{eqnarray}
{\cal S}&=&\int d^2 \sigma d^2 \theta\,\left(
2iV_{ \bar \mu \nu }\,D'X^{ \bar \mu }DX^ \nu -2iV_{ \mu \bar \nu }\,
D'X^ \mu DX^{ \bar \nu }+ V_ \mu \, \partial _ \sigma X^ \mu - V_{ \bar \mu } \,\partial _ \sigma
X^{ \bar \mu }
\right) \nonumber\\
&&\qquad\qquad+i\,\int d \tau d^2 \theta\, W.\label{actionB}
\end{eqnarray}
It is quite interesting to note that even here -- contrary to what is sometimes claimed -- the
theory remains invariant under K\"ahler transformations. Indeed, one readily verifies that,
\begin{eqnarray}
V(X, \bar X)&\rightarrow& V'( X, \bar X)= V(X, \bar X)+ f(X)+ \bar f( \bar X), \nonumber\\
W(X, \bar X)&\rightarrow& W'( X, \bar X)= W(X, \bar X)+i\left( f(X)- \bar f( \bar X)\right),
\end{eqnarray}
leaves the action eq.~(\ref{actionB}) invariant. Performing the integral over $\hat \theta $ in
eq.~(\ref{actionB}) yields eq.~(\ref{an12}) with vanishing torsion, $T=0$, and a K\"ahler metric
given by $g_{ \mu \bar \nu }=V_{ \mu \bar \nu }$. However, we find that eq.~(\ref{an12}) comes
with an extra, non-standard boundary term of the form\footnote{This unusual boundary term was
already noticed in \cite{Koerber:2003ef}. In order to recover the standard boundary term -- as
in eq.~(\ref{an122}), we will need non-trivial Neumann boundary conditions eq.~(\ref{nbc1}).},
\begin{eqnarray}
{\cal S}_{extra}=i\,\int d \tau d \theta \left( \left(V+i\,W\right)_{ \mu }D'X^{ \mu }+
\left(V-i\,W\right)_{ \bar  \mu }D'X^{ \bar \mu }\right).\label{caut1}
\end{eqnarray}
Varying the action in eq.~(\ref{actionB})\footnote{When varying we use the fact that the $N=2$
superfields $X^ \mu $ and $ X^{ \bar \mu }$ are unconstraind, while
$ \delta D'X^ \mu =-i\hat D \delta X^ \mu $ and similarly for $ \delta D' X^{ \bar \mu }$.},
yields besides the standard bulk equations of motion a boundary contribution given by,
\begin{eqnarray}
\delta {\cal S}\Big|_{boundary}=\int d \tau d^2 \theta \left(\left(V+i\,W\right)_ \mu \delta X^ \mu
-\left(V-i\,W\right)_{ \bar \mu } \delta X^{ \bar \mu }\right).\label{caut2}
\end{eqnarray}
Both eqs.~(\ref{caut1}) and (\ref{caut2}) indicate that the choice of boundary conditions will be
subtle here.

In order to get a feeling of what is going on, we first look at the simplest situation 
where there is only a single twisted chiral field (which we call $w$), {\em i.e.}~$n=1$. 
The model is characterized by two potentials $V(w, \bar w)$ 
and $W(w, \bar w)$. We get that the boundary term in the variation of the action, eq.~(\ref{caut2}), 
vanishes provided we impose the Dirichlet boundary condition,
\begin{eqnarray}
\delta  w= R^w{}_{ \bar w}\, \delta \bar w,\label{ssstr}
\end{eqnarray}
with,
\begin{eqnarray}
R^w{}_{ \bar w}\equiv\, \frac{V_{ \bar w}-i\, W_{ \bar w}}{V_w+i\,W_w}.
\end{eqnarray}
Eq.~(\ref{ssstr}) implies,
\begin{eqnarray}
\hat D  w= R^w{}_{ \bar w}\, \hat D \bar w,\label{ssstr2}
\end{eqnarray}
which using the constraints eq.~(\ref{n2tcsf}) reduces to the Neumann boundary condition,
\begin{eqnarray}
D'  w+ R^w{}_{ \bar w}\, D' \bar w=0.\label{ssstr3}
\end{eqnarray}
So the $ \sigma $-model describes open strings propagating on a K\"ahler manifold with K\"ahler 
potential $V$ in the 
presence of a D1-brane wrapped on a lagrangian submanifold (a trivial notion in two dimensions) 
whose position is determined by eq.~(\ref{ssstr}). In order to make contact with the example 
discussed at the end of section 2, 
we restrict ourselves to flat space, {\em i.e.} $V=( w+ \bar w)^2/2$, and 
assume that $W$ has the form $W=W(w+ \bar w)$. Using the coordinates defined in eq.~(\ref{newcds}), 
we identify $w=(Y^1+i\,Y^2)/\sqrt{2}$ and we find,
\begin{eqnarray}
W=-(w+ \bar w)\,Q'(w+ \bar w)+ Q( w+ \bar w),
\end{eqnarray}
where a prime denotes a derivative with respect to either $w$ or $ \bar w$ and $Q(w+ \bar w)$ is
a ``prepotential'' for $F$ which appears in eqs.~(\ref{zyz}) and (\ref{qqqs}),
\begin{eqnarray}
F= \partial _ w \partial _{ \bar w}\,Q(w + \bar w).
\end{eqnarray}
We get here,
\begin{eqnarray}
R^w{}_{ \bar w}= \frac{1+i\,Q''}{1-i\,Q''}= \frac{1+i\,F}{1-i\,F},
\end{eqnarray}
and this corresponds to the first choice ({\em i.e.}~the one for which $R_{ab}=R_{ba}$ holds) 
for $R^a{}_b$ made in section 2.
Using this we obtain the boundary conditions,
\begin{eqnarray}
-i( w- \bar w)-Q'(w+ \bar w)=\mbox{ constant},
\end{eqnarray}
and,
\begin{eqnarray}
D'w+D' \bar w=i\,Q''(w+ \bar w)\big(D'w-D' \bar w\big).
\end{eqnarray}
The resulting model is precisely the one discussed in section 2, however now in a manifest
$N=2$ supersymmetric setting. The extended supersymmetry fixed the choice of $R^a{}_b$.
Note that it is the potential $W$ which 
allows us to tune the precise location of the D1-brane. 

We now turn to the general case. The Dirichlet boundary conditions can be written as,
\begin{eqnarray}
\delta X^ \mu = R^ \mu {}_{ \bar \nu }\, \delta X^{ \bar \nu }+
R^ \mu {}_{  \nu } \,\delta X^{  \nu }.
\end{eqnarray}
Invariance of the boundary conditions under the supersymmetry transformations implies,
\begin{eqnarray}
\hat D X^ \mu = R^ \mu {}_{ \bar \nu }\, \hat D X^{ \bar \nu }+
R^ \mu {}_{  \nu }\, \hat D X^{  \nu },
\end{eqnarray}
which using the constraints eq.~(\ref{btcf}) results in,
\begin{eqnarray}
\left({\cal P}_+D'X\right)^ \mu =R^ \mu {}_ \nu\, D'X^ \nu .
\end{eqnarray}
Requiring this to be compatible with $ {\cal P}_+{\cal P}_+={\cal P}_+$ yields,
\begin{eqnarray}
&&R^ \mu{} _ \rho R^ \rho {}_ \nu = R^ \mu {}_ \nu , \nonumber\\
&&R^ \mu{} _{ \bar \rho }R^{ \bar \rho }{}_{ \bar \nu }=0.\label{ccc1}
\end{eqnarray}
Combining this with $R^a{}_cR^c{}_b= \delta ^ a_b$ gives in addition,
\begin{eqnarray}
&&R^ \mu {}_{ \bar \rho }R^{ \bar \rho }{}_ \nu = \delta ^ \mu _ \nu -R^ \mu {}_ \nu , \nonumber\\
&&R^ \mu{} _{  \rho }R^{  \rho }{}_{ \bar \nu }=0,\label{ccc2}
\end{eqnarray}
as well. Decomposing the complexified tangent space $T_{ {\cal M}}$ as
$ T_{ {\cal M}}=T^{(1,0)}_{ {\cal M}}\oplus T^{(0,1)}_{ {\cal M}}$, we see that eqs.~(\ref{ccc1})
and (\ref{ccc2}) imply the existence of projection operators
$ \pi _\pm: T^{(1,0)}_{ {\cal M}}\rightarrow T^{(1,0)}_{ {\cal M}}$,
\begin{eqnarray}
\pi _+^ \mu {}_ \nu \, \delta  X^ \nu &\equiv&R^ \mu {}_ \nu \, \delta  X^ \nu , \nonumber\\
\pi _-^ \mu {}_ \nu\, \delta  X^ \nu &\equiv &R^ \mu {}_{ \bar \rho }R^{ \bar \rho }{}_ \nu \, 
\delta  X^ \nu .
\end{eqnarray}
This allows us to rewrite the Dirichlet boundary conditions as,
\begin{eqnarray}
(\pi _- \delta X)^ \mu = R^ \mu {}_{ \bar \nu }\, \delta X^{ \bar \nu }.\label{lkj1}
\end{eqnarray}
For the Neumann directions we get,
\begin{eqnarray}
&&\left( \pi _+ {\cal P}_+D'X\right)^ \mu= R^ \mu {}_\nu D' X^ \nu , \nonumber\\
&&  \left( \pi _- {\cal P}_+D'X\right)^ \mu= 0.\label{lkj2}
\end{eqnarray}
Comparing this to eq.~(\ref{nbc1}) we conclude that we will have a non-degenerate magnetic
background in the $ \pi _+$ directions while the magnetic field vanishes in the $ \pi _-$
directions.

We first consider the case for which $\ker\, \pi _-=\emptyset$, {\em i.e.}~the only non-vanishing 
components of $R$ are $R^ \mu {}_ { \bar \nu } $ and $R^{ \bar \mu }{}_ \nu $. The Dirichlet 
boundary conditions are simply,
\begin{eqnarray}
\delta X^ \mu =R^ \mu {}_{ \bar \nu }\, \delta X^{ \bar \nu }.\label{dbsim}
\end{eqnarray}
This implies the integrability conditions,
\begin{eqnarray}
R^ \mu {}_{[ \bar \nu , \bar \rho ]}=R^ \sigma {}_{[ \bar \nu }R^ \mu {}_{ \bar \rho ], \sigma }.
\label{xqa}
\end{eqnarray} 
The boundary term in the variation of the action, eq.~(\ref{caut2}), will vanish provided,
\begin{eqnarray}
\left(V+i\,W\right)_ \mu \delta X^ \mu=
\left(V-i\,W\right)_{ \bar \mu } \delta X^{ \bar \mu },\label{vcxz}
\end{eqnarray}
which implies that,
\begin{eqnarray}
\left(V+i\,W\right)_ \mu R^ \mu{}_{ \bar \nu }=
\left(V-i\,W\right)_{ \bar \nu } ,
\end{eqnarray}
should hold. As a consequence, we find that besides eq.~(\ref{vcxz}), 
$ \left(V+i\,W\right)_ \mu D X^ \mu=
\left(V-i\,W\right)_{ \bar \mu } D X^{ \bar \mu }$ and
$\left(V+i\,W\right)_ \mu \dot X^ \mu=
\left(V-i\,W\right)_{ \bar \mu } \dot X^{ \bar \mu }$ hold as well. Using 
$D^2=-(i/2) \partial _ \tau $ we get that the previous is consistent provided,
\begin{eqnarray}
V_{  \mu \bar \rho }\,R^ { \bar \rho} {}_{  \nu }=
V_{ \nu \bar \rho }\,R^{ \bar  \rho} {}_{  \mu },\label{qpo}
\end{eqnarray}
holds, {\em i.e.}~$R_{ \mu \nu }= R_{ \nu \mu }$. 
Introducing a set of real worldvolume coordinates $ \sigma ^ \tau $, 
$ \tau \in\{ 1,\cdots n\}$, we get that eq.~(\ref{xqa}) guarantees that,
\begin{eqnarray}
\frac{ \partial X^ \mu }{ \partial \sigma ^ \tau }=R^ \mu {}_{ \bar \nu }
\frac{ \partial X^{ \bar \nu }}{ \partial \sigma ^ \tau },
\end{eqnarray}
is satisfied. With this and eq.~(\ref{qpo}), one finds immediately that the pullback of the K\"ahler 
two-form to the worldvolume of the brane vanishes.
This shows that, whenever $\ker\, \pi _-=\emptyset$, we have a D$n$-brane which wraps an
isotropic submanifold of maximal dimension, {\em i.e.}~a lagrangian
submanifold.\footnote{For the definition of isotropic and lagrangian
submanifolds, see appendix \ref{app subm}.} 
{}From eq.~(\ref{dbsim}), we get that $ \hat D X^ \mu =R^ \mu {}_{ \bar \nu }\, \hat D 
X^{ \bar \nu }$, which using the constraints gives the Neumann boundary conditions,
\begin{eqnarray}
D' X^ \mu +R^ \mu {}_{ \bar \nu }\,D' X^{ \bar \nu }=0,
\end{eqnarray}
from which it
follows that for a lagrangian D-brane the magnetic field is
necessarily zero. In other words, a lagrangian D-brane can only
carry a line bundle with flat connection.

We now come to the case where $\ker\, \pi _-\neq\emptyset$. In order to proceed, we
assume the existence of adapted coordinates $X^{\check \mu }$ and $X^{\hat \mu }$
(and their complex conjugates), $\check \mu , \check \nu , \cdots\in\{1,\cdots , k\}$ and
$\hat \mu , \hat \nu , \cdots\in\{k+1,\cdots , n\}$, such that the only non-vanishing components
of $ \pi _+$ and $ \pi _-$ are 
$ \pi _-^{\hat \mu }{}_{ \hat \nu }= \delta ^{ \hat \mu }_{ \hat \nu }$
and $ \pi_+^{\check \mu }{}_{\check \nu }= \delta ^{ \check \mu }_{\check \nu }$. The only 
non-vanishing components of $R$ are then $R^{ \hat \mu }{}_{ \bar { \hat \nu }}(\hat X,\check X)$
and $R^{\check \mu }{}_{\check \nu }= \delta ^{\check \mu }_{\check \nu} $.
The Dirichlet boundary conditions become,
\begin{eqnarray}
\delta X^{ \hat \mu }=R^{ \hat \mu }{}_{ \bar{ \hat \nu }} \delta X^{\bar {\hat \nu} }.\label{njer}
\end{eqnarray}
The resulting integrability conditions imply that $R^{ \hat \mu }{}_{ \bar{ \hat \nu }}$ does not 
depend on $X^{\check \mu }$ or $ X^{\bar{\check \mu }}$ (so $R^{ \hat \mu }{}_{ \bar { \hat \nu 
}}=R^{ \hat \mu }{}_{ \bar { \hat \nu }}(\hat X)$) and,
\begin{eqnarray}
R^{\hat \mu} {}_{[ \bar{\hat \nu} , \bar{\hat \rho} ]}=R^ {\hat\sigma} 
{}_{[ \bar {\hat\nu} }R^ {\hat\mu} {}_{ \bar {\hat\rho }], {\hat\sigma} }.
\end{eqnarray}
A necessary -- but not sufficient -- condition for the vanishing of the boundary term in
eq.~(\ref{caut2}) is that,
\begin{eqnarray}
\left(V+i\,W\right)_ {\hat\mu }\delta X^{\hat \mu}=
\left(V-i\,W\right)_{ \bar {\hat\mu} } \delta X^{ \bar {\hat\mu }},\label{vcxz2}
\end{eqnarray}
which requires that,
\begin{eqnarray}
\left(V+i\,W\right)_ {\hat\mu} R^ {\hat\mu}{}_{ \bar {\hat\nu} }=
\left(V-i\,W\right)_{ \bar{\hat \nu} } ,
\end{eqnarray}
should hold.
Eq.~(\ref{vcxz2}) also implies that,
\begin{eqnarray}
V_{ \hat \mu \bar {\hat \rho }}R^{ \bar {\hat \rho }}{}_{\hat \nu }=
V_{ \hat \nu \bar {\hat \rho }}R^{ \bar {\hat \rho }}{}_{\hat \mu },
\end{eqnarray}
or $R_{ \hat \mu \hat \nu }= R_{ \hat \nu \hat \mu }$.
{}From eq.~(\ref{njer}) and the bulk constraints eq.~(\ref{btcf}) we obtain part of the Neumann 
boundary conditions,
\begin{eqnarray}
D'\, X^{ \hat \mu }+R^{ \hat \mu }{}_{ \bar{ \hat \nu }} D'\, X^{\bar {\hat \nu} }=0.\label{njer6}
\end{eqnarray}

With this the boundary term in the variation of the action, eq.~(\ref{caut2})
does not vanish yet.  Denoting the coordinates $X^{\check \mu }$ and $X^{ \bar {\check \mu }}$ 
collectively by
$X^{\check a}$ and introducing the canonical complex structure
$J^{ \check a}{}_{ \check b}$\footnote{Its
nonvanishing components are
$J^{\check \mu }{}_{\check \nu }=+i\, \delta ^{\check \mu }_{\check \nu}$ and
$J^{ \bar {\check \mu} }{}_{ \bar {\check \nu} }=-i\,
\delta ^{ \bar {\check \mu }}_{ \bar {\check \nu} }$.}, we rewrite eq.~(\ref{caut2}) using
eq.~(\ref{vcxz2}),
\begin{eqnarray}
\delta {\cal S}\Big|_{boundary}=-i\,\int d \tau d^2 \theta \left(
V_{\check b}J^{\check b}{}_{\check a}-W_{\check a}\right)\delta X^{\check  a}.\label{caut22}
\end{eqnarray}
Eq.~(\ref{lkj2}) suggested the presence of a non-degenerate magnetic field $F_{\check a\check b}$
which implies Neumann boundary conditions of the form,
\begin{eqnarray}
D'X^{\check a}=F^{\check a}{}_{ b}\,DX^{ b},\label{nxc1}
\end{eqnarray}
where indices without checks or hats run from 1 through $d=2n$.
Using the fact that the bulk constraints eq.~(\ref{btcf}) can be
rewritten as,
\begin{eqnarray}
\hat D X^{\check a}=J^{\check a}{}_{\check b}\,D'X^{\check
b},\label{nxc2}
\end{eqnarray}
we propose Neumann boundary conditions of the form,
\begin{eqnarray}
\hat D X^{\check a}=K^{\check a}{}_b\,DX^b,\label{nxc3}
\end{eqnarray}
with $K^{\check a}{}_b=J^{\check a}{}_{\check c}F^{\check c}{}_b$.
Combining $\hat D^2=-(i/2) \partial _ \tau $ with eq.~(\ref{nxc3}) we get that eq.~(\ref{nxc3})
must be of the form,
\begin{eqnarray}
\hat D X^{\check a}=K^{\check a}{}_{\check b}DX^{\check b},\label{nxc4}
\end{eqnarray}
and $K^{\check a}{}_{\check b}$ is a complex structure ({\em i.e.}~it squares to $-1$ and its Nijenhuis
tensor vanishes) which depends only on $X^{\check a}$. This explains in a natural way the emergence
of an extra complex structure when dealing with coisotropic branes \cite{Kapustin:2001ij},
\cite{Kapustin:2003kt}, \cite{Lindstrom:2002jb}: imposing constraints linear in the fermionic
derivatives does give rise to complex structures.

When analyzing the boundary term in the variation of the $N=2$ action,
eq.~(\ref{caut22}),
one has to take into account that $\hat X$ is constrained by eq.~(\ref{nxc4}). As
a consequence we have that,
\begin{eqnarray}
\delta X^{\check a}= \frac{ \partial X^{\check a}}{ \partial \tilde{X}^{ \check  b}}\left(
\hat D \delta \Lambda ^{\check  b}- \tilde K ^{\check  b}{}_{\check  c}D \delta \Lambda ^{\check  c}
\right),
\end{eqnarray}
where $ \tilde X$ are coordinates in which the complex structure $ K$ is constant (which we denote
by $ \tilde K$) and we expressed $ \tilde X$ in terms of unconstrained fermionic superfields
$ \Lambda$:
$\tilde{X}^{\check a}=\hat D\, \Lambda ^{\check a}-\tilde{K}^{\check a}{}_{\check b}\,
D \Lambda ^{\check b}$. Using
this in eq.~(\ref{caut22}), we find that it becomes,
\begin{eqnarray}
\delta {\cal S}\Big|_{boundary}&=&-i\,\int d \tau d^2 \theta\, \delta  \Lambda ^{\check e}
DX^{\check b}
\frac{ \partial X^{\check a}}{ \partial \tilde X^{\check  e}}\Big(
2M_{\check c}K^{\check c}{}_{[\check b,\check a]}+M_{\check a,\check c}
K^{\check c}{}_{\check b}-M_{\check c,\check b}K^{\check c}{}_{\check a}\Big) \nonumber\\
&&-i\,\int d \tau d^2 \theta\, \delta  \Lambda ^{\check e}
\frac{ \partial X^{\check a}}{ \partial \tilde X^{\check  e}}
\Big(M_{\check a, \hat b}\hat DX^{ \hat b}- M_{\check c,\hat b}K^{\check c}{}_{\check a}
DX^{ \hat b}\Big)
, \label{bbb2}
\end{eqnarray}
where,
\begin{eqnarray}
M_{\check a}\equiv V_{\check b} J^{\check b}{}_{\check a}-W_{\check a},
\end{eqnarray}
and where we denoted the coordinates $X^{ \hat \mu }$ and $ X^ {\bar { \hat \mu }}$ collectively by
$X^{ \hat a}$.
Using eq.~(\ref{njer}) and the fact that 
$R^{\hat \mu }{}_{ \bar { \hat \nu }}$ does not depend on $\check X$, one shows that the second 
line in eq.~(\ref{bbb2}) vanishes provided,
\begin{eqnarray}
V_{ \hat \mu \bar {\check \nu }}= V_{\check \mu \bar { \hat \nu }}=0,
\end{eqnarray}
{\em i.e.}~the K\"ahler potential factorizes (modulo a K\"ahler transformation) as
$V= \hat V( \hat X, \bar { \hat X})+\check V(\check X, \bar {\check X})$.
We rewrite the argument of the first line in eq.~(\ref{bbb2}) as,
\begin{eqnarray}
&&2M_{\check c}K^{\check c}{}_{[\check b,\check a]}+M_{\check a,\check c}
K^{\check c}{}_{\check b}-M_{\check c,\check b}K^{\check c}{}_{\check a}= \nonumber\\
&&\qquad\qquad
2F_{\check a\check b}+ \partial _{\check a}
\big( V_{\check c}(JK)^{\check c}{}_{\check b}-W_{\check c}K^{\check c}{}_{\check b}
\big)-\partial _{\check b}
\big( V_{\check c}(JK)^{\check c}{}_{\check a}-W_{\check c}K^{\check c}{}_{\check a}
\big),
\end{eqnarray}
where,
\begin{eqnarray}
F_{\check a\check b}\equiv -\omega _{\check a\check c}\,K^{\check c}{}_{\check b}=
-g _{\check a\check c}\,\big(JK\big)^{\check c}{}_{\check b},\label{defF}
\end{eqnarray}
with $ \omega $ the K\"ahler form ($ \omega _{ab}\equiv g_{ac}J^c{}_b$).
{}From this we read that the boundary term in the variation, eq.~(\ref{bbb2}), does vanish provided
that $F_{ab}$ is a closed 2-form. Locally we get that
\begin{eqnarray}
F_{\check a\check b}= \partial _{\check a}A_{\check b}-\partial _{\check b}A_{\check a},\label{eee4}
\end{eqnarray}
with,
\begin{eqnarray}
A_{\check a}= - \frac{1}{2}V_{\check c}(JK)^{\check c}{}_{\check a}+
\frac{1}{2}W_{\check c}K^{\check c}{}_{\check a}+ \partial _{\check a}f,\label{eee5}
\end{eqnarray}
with $f$ an arbitrary real function. Given $F$, eqs.~(\ref{eee4}) and (\ref{eee5}) constrain the
potential $W$. From the fact that $F_{\check a\check b}$ is antisymmetric in its indices we
immediately find that both $F_{\check a\check b}$ and $ \omega _{\check a\check b}$ are
$(2,0)+(0,2)$ forms with respect to the complex structure $K$ implying -- as
$ \omega $ is non-degenerate -- that $k=2l$, $l\in\IN$. As a consequence the dimension of the
submanifold spanned by $X^{\check a}$ is a multiple of four \cite{Kapustin:2001ij},
\cite{Kapustin:2003kt}, \cite{Lindstrom:2002jb}. So here we are dealing with open strings
in the presence of a coisotropic D$(n+2l)$-brane. An obvious realization of the previous
is given by the case in which the submanifold parametrized by the coordinates $X^{\check a}$ is
hyper-K\"ahler.

The reason for calling these D-branes coisotropic is that they wrap coisotropic submanifolds. 
Denoting the submanifold wrapped by the D-brane by ${\cal N} \subset {\cal M}$, each tangent space 
of ${\cal N}$ is a subspace of the tangent space of ${\cal M}$, $T_{\cal N} \subset T_{\cal M}$. For 
${\cal N}$ to be coisotropic, we need $T_{\cal N}^\bot \subset T_{\cal N}$, where $T_{\cal N}^\bot$ 
is the symplectic complement of $T_{\cal N}$.\footnote{Again, see appendix \ref{app subm} for 
definitions.} The complement $T_{\cal N}^\bot$ is generated by those tangent  
vectors along the brane which are in the image of $\pi_-$. We will  
denote them by $\delta \sigma^{\hat a}$. These are symplectic  
orthogonal to themselves because of the relation eq.~(\ref{njer}) and  the  
symmetry of $R$. On the other hand they are symplectic orthogonal to  
all vectors in $\mbox{Im}\,\pi_+$, because the factorization of the  
metric implies $\omega_{\hat\mu \bar{\check\nu}} = \omega_{\check\mu  
\bar{\hat\nu}} = 0$. No other vectors of $\mbox{Im}\,\pi_-$ can be  
orthogonal to the $\sigma^{\hat a}$, because $\omega$ is non- 
degenerate. This shows that indeed $T_{\cal N}^\bot = \{\delta 
\sigma^{\hat a}\} \subset T_{\cal N} =  \{\delta\sigma^{\hat a},  
\delta X^{\check a}\}$.
Whenever $k=0$, we find that $T_{\cal N}^\bot = T_{\cal N}$, so that, as mentioned before, $
{\cal N}$ becomes lagrangian. In the other extreme, when $k=n$, we find a maximally coisotropic 
D(4l)-brane wrapping the entire target space ${\cal M}$. This is obviously only possible for target 
space dimensions which are a multiple of four. In general, the magnetic flux $F$, the pullback of 
$\omega$ and the additional complex structure $K = -\omega^{-1} F$ are only nonzero on the 
$4l$-dimensional quotient space $T_{\cal N}/T_{\cal N}^\bot = \{\delta X^{\check a}\}$, where they 
are all non-degenerate.

Upon using eq.~(\ref{njer6}) we can rewrite the non-standard boundary term in the $N=1$ action,
eq.~(\ref{caut1}), as,
\begin{eqnarray}
{\cal S}_{extra}=i\,\int d \tau d \theta \left(V_{\check a}+W_{\check b}
J^{\check b}{}_{\check a}\right)D'X^{\check a}.\label{nsccc}
\end{eqnarray}
Using the boundary condition eq.~(\ref{nxc4}) and eq.~(\ref{eee5})
this indeed reduces to the standard boundary term eq.~(\ref{an122}).

We will finish this section with an example for $n=2$ (or $d=4$). We have two twisted chiral fields
which we denote by $z$ and $w$. The K\"ahler potential is of the form $V(z- \bar z, w+ \bar w)$.
Imposing Dirichlet boundary conditions,
\begin{eqnarray}
\mbox{Re}\, z=\mbox{ constant},\qquad \mbox{Im}\, w=\mbox{ constant},
\end{eqnarray}
and Neumann boundary conditions,
\begin{eqnarray}
D'\,\mbox{Im}\, z=D'\,\mbox{Re}\, w=0,
\end{eqnarray}
we find that the action,
\begin{eqnarray}
{\cal S}=\int d^2 \sigma d^2 \theta D'\hat D' \,V(z- \bar z, w+ \bar w)\,,
\end{eqnarray}
describes open strings propagating on a K\"ahler manifold in
the presence of a D2-brane wrapped around a lagrangian submanifold.

{From} the previous discussion we know that there exists the
possibility of a (maximally) coisotropic brane -- in the present
case a D4-brane -- as well. This can certainly be (locally) realized
if the K\"ahler potential $V(z- \bar z, w+ \bar w)$ is actually
hyper-K\"ahler, which is indeed so if the potential satisfies the
Monge-Amp\`ere equation,
\begin{eqnarray}
V_{z \bar z}\,V_{w \bar w}-V_{z \bar w}\,V_{w \bar z}=1.\label{MAeq}
\end{eqnarray}
The Legendre transform method \cite{Hitchin:1986ea} allows us to construct $V$ in terms of a complex
prepotential $h(x+z- \bar z)$ with $x\in \IR$. The K\"ahler potential is then given by the following
Legendre transform,
\begin{eqnarray}
V(z- \bar z, w+ \bar w)=h(x+z- \bar z)+ \bar h(x+ \bar z-z)-x\,(w+ \bar w).\label{exLT}
\end{eqnarray}
Flat space corresponds {\em e.g.}~to,
\begin{eqnarray}
h=- \frac{1}{4}\,\left(x+z- \bar z \right)^2.\label{fsex}
\end{eqnarray}
The metric can be expressed in terms of the prepotential,
\begin{eqnarray}
g_{z \bar z}&=&V_{z \bar z}=-4\, \frac{h'' \bar h''}{h''+ \bar h'' }\,, \nonumber\\
g_{z \bar w}&=&V_{z \bar w}= \frac{ h''- \bar h''}{h''+ \bar h''}\,, \nonumber\\
g_{w \bar z}&=&V_{w \bar z}= \frac{ \bar  h''-h''}{h''+ \bar h''}\,,\nonumber\\
g_{w \bar w}&=&V_{w \bar w}=- \frac{1}{h''+ \bar h''}\, ,\label{exmet}
\end{eqnarray}
where $h''\equiv \partial _x^2h(x+z- \bar z)$ and similarly for $ \bar h''$.
The complex structure $K$ is given by,
\begin{eqnarray}
K^{z}{}_{ \bar z}=g_{w \bar z},\quad
K^{z}{}_{ \bar w}=g_{w \bar w},\quad
K^{w}{}_{ \bar z}=-g_{z \bar z},\quad
K^{w}{}_{ \bar w}=-g_{z \bar w},
\end{eqnarray}
which, upon using eq.~(\ref{defF}) and (\ref{exmet}), gives the
magnetic background,
\begin{eqnarray}
F_{zw}=+i,\qquad F_{ \bar z \bar w}=-i.
\end{eqnarray}
Using coordinates in which $K$ is constant,
\begin{eqnarray}
r\equiv z+ \bar z,\quad s\equiv V_w, \quad t\equiv i( \bar w-w),\quad u\equiv iV_z,
\end{eqnarray}
one easily determines $W$ such that the boundary term in the variation of the action
eq.~(\ref{caut2}) vanishes,
\begin{eqnarray}
W= \frac{i}{2}\left(z\,V_z+w\,V_w- \bar z\,V_{ \bar z}- \bar w\, V_{ \bar w}\right).\label{exWex}
\end{eqnarray}
So the action eq.~(\ref{aaa0}) with $V$ given by eq.~(\ref{exLT}) and $W$ by eq.~(\ref{exWex})
together with the Neumann boundary conditions,
\begin{eqnarray}
\hat Dz=V_{w \bar z}\,D \bar z+V_{w \bar w}\,D \bar w\,,\qquad
\hat Dw=-V_{z \bar z}\,D \bar z-V_{z \bar w}\,D \bar w\,,\label{exWexN}
\end{eqnarray}
describes open strings in the presence of a maximally coisotropic D4-brane.
Taking flat space eq.~(\ref{fsex}), one recovers {\em e.g.}~the example studied in
\cite{Font:2006na}.

\section{Type B branes}
We start from the most general $N=2$ invariant action,
\begin{eqnarray}
{\cal S}=-\int d^2 \sigma d^ 2 \theta D'\hat D'\,V(X, \bar X)+ i\, \int d \tau d ^2 \theta
\,W(X, \bar X),\label{aatypb}
\end{eqnarray}
where $V$ and $W$ are real scalar functions of the chiral superfields $X$ and $ \bar X$ which were
defined in eq.~(\ref{bcf}). Working out the $D'$ and $\hat D'$ derivatives we get,
\begin{eqnarray}
{\cal S}= -2i\int d^2 \sigma d^ 2 \theta \, V_{ \alpha \bar \beta
}\, \left( DX^ \alpha DX^{ \bar \beta }- D'X^ \alpha D'X^{ \bar
\beta }\right) +i\, \int d \tau  d^2 \theta \,W(X, \bar
X).\label{atyb}
\end{eqnarray}
Note that even in the presence of boundaries, the action remains invariant under K\"ahler
transformations,
\begin{eqnarray}
V(X, \bar X)&\rightarrow& V'( X, \bar X)= V(X, \bar X)+ f(X)+ \bar f( \bar X).
\end{eqnarray}
In addition we have the following invariance as well,
\begin{eqnarray}
W(X, \bar X)&\rightarrow& W'( X, \bar X)= W(X, \bar X)+ g(X)+ \bar
g( \bar X).
\end{eqnarray}

Performing the integral over $ \hat \theta $ and comparing the
result to the $N=1$ action in eq.~(\ref{an12}), we find that the
target space is a K\"ahler manifold with K\"ahler potential $V(X,
\bar X)$ -- {\em i.e.} the non-vanishing components of the metric
are $g_{ \alpha \bar \beta }=V_{ \alpha \bar \beta }$ -- which
carries a $U(1)$ bundle where the non-vanishing components of the
magnetic field $F_{ab}\equiv b_{ab}$ are determined by the potential
$W( X, \bar X)$,
\begin{eqnarray}
F_{ \alpha \bar \beta }= - F_{ \bar \beta \alpha }=-i\,W_{ \alpha \bar \beta },\qquad
F_{ \alpha \beta }= F_{ \bar a \bar \beta }=0.
\end{eqnarray}
The last equation states that we are dealing with a holomorphic vector bundle.

When varying the action eqs.~(\ref{aatypb}) or (\ref{atyb}), one
needs to take the constraints eqs.~(\ref{bcf}) or (\ref{n2csf}) into
account. E.g. working in $N=(2,2)$ superspace, we
express\footnote{Once more, for conventions we refer to the
appendix.} $X$ in terms of an unconstrained superfield $L$: $X^
\alpha = \bar \ID_+ \bar \ID_- L^ \alpha = 2\, \bar \ID'\, \bar
\ID\,L^ \alpha $. In $N=2$ superspace one has unconstrained $N=2$
fields $ \Lambda ^ \alpha $, $ \Lambda ^{ \bar \alpha }$ and $M^
\alpha $, $M^{ \bar \alpha }$, in terms of which we get,
\begin{eqnarray}
&&X^ \alpha =\left(\hat D-iD\right) \Lambda ^ \alpha ,\qquad
X^{ \bar \alpha }=\left(\hat D+i D\right) \Lambda ^{ \bar \alpha }, \nonumber\\
&& D'X^ \alpha =\left(\hat D-iD\right) M ^ \alpha-
\partial _ \sigma \Lambda ^ \alpha  ,\qquad
D'X^{ \bar \alpha }=\left(\hat D+i D\right) M ^{ \bar \alpha }+
\partial_ \sigma  \Lambda ^{ \bar \alpha }.\label{solcon}
\end{eqnarray}
Using this we get the boundary term in the variation of the action eq.~(\ref{aatypb}) or
(\ref{atyb}),
\begin{eqnarray}
\delta {\cal S}\Big|_{boundary}&=&-2i\,\int d \tau d^2 \theta \bigg(
\delta \Lambda ^ { \alpha } \left(V_{ \alpha \bar \beta }D'X^{ \bar \beta }+ i
W_{ \alpha \bar \beta} DX^{ \bar \beta }\right) \nonumber\\
&&\qquad+\delta \Lambda ^ { \bar  \alpha } \left(V_{ \bar  \alpha \beta }D'X^{ \beta }- i
W_{ \bar \alpha \beta} DX^{ \beta }\right)
\bigg).\label{bbb9}
\end{eqnarray}
Once again we need suitable boundary conditions to cancel this. We impose Dirichlet boundary
conditions on the unconstrained $N=2$ superfields $ \Lambda $,
\begin{eqnarray}
\delta \Lambda ^ \alpha = R^ \alpha {}_ \beta\, \delta \Lambda ^ \beta +
R^ \alpha {}_{ \bar \beta } \,\delta \Lambda ^{ \bar \beta }.
\end{eqnarray}
As $(\hat D-iD) \Lambda ^{ \bar \alpha }$ should not appear in $
\delta X^ \alpha $, we necessarily need that,
\begin{eqnarray}
R^ \alpha {}_{ \bar \beta }=R^ { \bar \alpha } {}_{ \beta }=0.
\end{eqnarray}
We find that $ \delta X^ \alpha =R^ \alpha {}_ \beta \,\delta X^ \beta $ follows from
$ \delta \Lambda ^ \alpha =R^ \alpha {}_ \beta\, \delta \Lambda ^ \beta $ provided,
\begin{eqnarray}
R^ \alpha {}_{ \delta  , \bar \epsilon  }\, {\cal P}_+^ \delta  {}_ \beta\,
{\cal P}_+^{ \bar \epsilon }{}_{ \bar \gamma }=0,\label{bic1}
\end{eqnarray}
is satisfied. Finally, requiring that $DX^ \alpha = {\cal P_+}^ \alpha{} _ \beta\, D X^ \beta  $ and
$ \partial _ \tau X^ \alpha = {\cal P}_+^ \alpha {}_ \beta\, \partial _ \tau X^ \beta $
are mutually compatible gives the condition,
\begin{eqnarray}
R^ \alpha {}_{ \delta , \epsilon } \,{\cal P_+}^ \delta {}_{[ \beta }\,
{\cal P_+}^ \epsilon  {}_{ \gamma ] }=0.\label{bic2}
\end{eqnarray}
Eqs.~(\ref{bic1}) and (\ref{bic2}) guarantee the existence of coordinates
$X^{\hat \alpha } $, $ \hat\alpha \in\{k+1,\cdots m\}$ 
 where $k$ is the rank of $ {\cal P}_+$, such that the Dirichlet
boundary conditions are given by,
\begin{eqnarray}
X^{\hat \alpha }=\mbox{ constant}.
\end{eqnarray}
Denoting the remainder of the coordinates by 
$X^{\tilde \alpha }$, $\tilde \alpha \in\{1,\cdots, k\}$, we find that
eq.~(\ref{bbb9}) vanishes provided we impose the Neumann boundary conditions,
\begin{eqnarray}
V_{ \tilde \alpha \bar \beta }\,D'X^{ \bar \beta }=-i\,W_{ \tilde \alpha \bar { \tilde \beta }}
\,DX^{ \bar { \tilde \beta }},
\end{eqnarray}
where $ \bar \beta $ runs from $1$ through $m$.

In this situation the $ \sigma $-model describes open strings in a
background with a D$2k$-brane wrapped on a holomorphic
submanifold. In addition, the D-brane can carry non-trivial magnetic
flux as long as this corresponds to the curvature of a connection on
a holomorphic line bundle. Note that, in contrast to the A brane
case, the conditions on the $U(1)$ flux are independent of the
geometry of the brane.

\section{Duality transformations}\label{sec:dual}
\subsection{Generalities}
Supersymmetric non-linear $ \sigma $-models allow for various
duality transformations interchanging the different types of
superfields \cite{Gates:1984nk}, \cite{Rocek:1991vk},
\cite{Grisaru:1997ep}, \cite{Bogaerts:1999jc},
\cite{Lindstrom:2007sq}, \cite{Merrell:2007sr}. Here we are chiefly
interested in duality transformations interchanging chiral and
twisted chiral fields and vice-versa. Let us first briefly review
the case without boundaries. The basic idea is to start with a
potential with an isometry. Subsequently one gauges the isometry and
imposes -- using Lagrange multipliers -- that the gauge fields are
pure gauge. Integrating over the Lagrange multipliers gives back the
original model while integrating over the gauge fields (or their
potentials which are unconstrained superfields) yields the dual
model.

We start from the $N=(2,2)$ action (without boundaries),
\begin{eqnarray}
{\cal S}=\int d^2 \sigma \,d^4 \theta \left(
-\int dy\, W(y,\cdots)+(z+ \bar z)\,y
\right),\label{fstor}
\end{eqnarray}
where $y$ is an unconstrained $N=(2,2)$ superfield, $z$ is either a chiral or a twisted chiral
superfield and $\cdots$ stand for other, spectator fields. The equations of motion for
$y$ give,
\begin{eqnarray}
z+ \bar z=W(y,\cdots ),
\end{eqnarray}
which upon inversion gives,
\begin{eqnarray}
y=U( z+ \bar z,\cdots).
\end{eqnarray}
Using this to eliminate $y$ yields the second order action,
\begin{eqnarray}
{\cal S}= \int d^2 \sigma\, d^4 \theta\,\int d(z+ \bar z)\,U(z+ \bar z,\cdots).
\end{eqnarray}
When however taking $z$ and $ \bar z$ to be chiral and integrating over them in
eq.~(\ref{fstor}) we get,
\begin{eqnarray}
\bar \ID_+ \bar \ID_-y=\ID_+\ID_-y=0,
\end{eqnarray}
which is solved by putting $y= w+ \bar w$ with $w$ a twisted chiral superfield.
If on the other hand we started off with a field $z$ which was twisted chiral we
get upon integrating over $z$ and $ \bar z$,
\begin{eqnarray}
\bar \ID_+  \ID_-y=\ID_+\bar\ID_-y=0,
\end{eqnarray}
which is now solved by putting $y = w + \bar w$, with $w$ a chiral superfield.
The resulting second order action (which is the action one starts with)
is in both cases given by,
\begin{eqnarray}
{\cal S}=-\int d^2 \sigma \,d^4 \theta \,\int d(w+ \bar w)\, W(w+ \bar w,\cdots).
\end{eqnarray}
So we conclude that this duality
transformation -- associated with a $U(1)$ isometry -- allows one to exchange chiral for twisted
chiral fields and vice-versa. The natural question which arises here is whether this duality
symmetry persists when boundaries are present. The main difficulty will be to introduce the right 
boundary terms such that the boundary conditions of the various fields remain consistent with the
duality transformation.

\subsection{From B to A branes}
We start our investigation with B-branes which are fairly well under control. The initial model
has $n$ chiral fields $z^ \alpha $, $ \alpha \in\{1,\cdots,n\}$ 
and it is characterized by a K\"ahler potential $V(z+ \bar z)$ and a $U(1)$ prepotential 
$W( z+ \bar z)$. As the notation already indicates the potentials are such that 
$ \partial _ \alpha V= \partial _{ \bar \alpha }V$ and 
$ \partial _ \alpha W= \partial _{ \bar \alpha }W$ hold, implying the existence of $n$ isometries
which should allow us to dualize the model to an A-brane. The action is given by eq.~(\ref{aatypb})
and we choose the boundary conditions as fully Neumann,
\begin{eqnarray}
V_{ \bar \alpha \beta }(z+ \bar z)\,D'z^{ \beta }&=&+i W_{ \bar \alpha  \beta }(z+ \bar z)
\,Dz^{ \beta }, \nonumber\\
V_{ \alpha \bar \beta }(z+ \bar z)\,D'z^{ \bar \beta }&=&-i W_{ \alpha \bar \beta }(z+ \bar z)
\,Dz^{ \bar \beta }.\label{bdy41}
\end{eqnarray}
We introduce a set of unconstrained real superfields $y^ \alpha =(y^{ \alpha })^\dagger$ (the
gauge fields which in a second order formulation of the model will be identified with 
$z^ \alpha + z^{ \bar \alpha }$) which satisfy the boundary conditions,
\begin{eqnarray}
V_{ \alpha  \beta }(y)\,D'y^{ \beta }&=&+W_{ \alpha \beta }(y)\,\hat D y^{ \beta }, \nonumber\\
V_{ \alpha  \beta }(y) \hat D'y^{ \beta }&=&-W_{ \alpha \beta }(y)\, D y^{ \beta },
\label{bdygf}
\end{eqnarray}
where we used the isometries of $V$ and $W$. The first order action is given by,
\begin{eqnarray}
{\cal S}&=&-\int d^2 \sigma d^ 2 \theta\Big( D'\hat D'\, V(y)
-2i\,w_ \alpha \,\ID_- \bar \ID_+\,y^ \alpha 
-2i\,w_{ \bar \alpha } \,\ID_+ \bar \ID_-\,y^ \alpha\Big) \nonumber\\
&&\qquad 
+ \,i\, \int d \tau d ^2 \theta\,\bigg( W(y)-y^ \alpha  \,
\frac{ \partial W(y)}{ \partial y^ \alpha }\Big|_{y=y( w+ \bar w) }
-i\,y^ \alpha \,\big(w_ \alpha -w_{ \bar \alpha }\big)\bigg)
\nonumber\\
&=&\int d^2 \sigma d^ 2 \theta D'\hat D'\,\Big( -V(y)
+ y^ \alpha \big(w_ \alpha +  w_{ \bar \alpha }\big)\Big) \nonumber\\
&&\qquad +\,i\, \int d \tau d ^2 \theta
\,\bigg(W(y)-  y^ \alpha\,\frac{ \partial W(y)}{ \partial y^ \alpha }\Big|_{y=y(w+ \bar w) }\bigg),
\label{fo80}
\end{eqnarray}
where the two forms of the action are related through partial integration and use of the 
constraints. When writing $y=y( w+ \bar w)$, we mean that the $ y^ \alpha $'s are given as a 
function of the $w_ \alpha + w_{ \bar \alpha }$'s such that,
\begin{eqnarray}
\frac{ \partial V(y)}{ \partial y^ \alpha }= w_ \alpha + w_{ \bar a},\label{bulkA}
\end{eqnarray}
holds.
In the first expression for the action, $w_ \alpha $ and $ w_{ \bar \alpha }$ are 
unconstrained $N=2$ superfields while in the second form for the action they are $N=(2,2)$
twisted chiral superfields. 

Varying $w_ \alpha $ and $ w_{ \bar \alpha }$ in the first 
form of the action
gives the bulk equation of motion,
\begin{eqnarray}
\ID_- \bar \ID_+\,y\Big|_{ \theta '= \hat \theta '=0}=
\ID_+ \bar \ID_-\,y\Big|_{ \theta '= \hat \theta '=0}=0.\label{cfs1}
\end{eqnarray}
These constraints are themselves twisted chiral fields implying (by acting with $D$ and $\hat D$ on 
them) that eq.~(\ref{cfs1}) is equivalent to the full $N=(2,2)$ superspace constraints,
\begin{eqnarray}
\ID_- \bar \ID_+\,y=\ID_+ \bar \ID_-\,y=0,
\end{eqnarray}
which are solved by putting,
\begin{eqnarray}
y^ \alpha = z^ \alpha + z^{ \bar \alpha },\label{zbzex}
\end{eqnarray}
with $z^ \alpha $ chiral superfields. The variation yields a boundary term as well which vanishes if 
we impose the Dirichlet boundary conditions on the Lagrange multipliers,
\begin{eqnarray}
-i\big(w_ \alpha -w_{ \bar \alpha }\big)-
\frac{ \partial W(y)}{ \partial y^ \alpha }\Big|_{y=y( w+ \bar w) }=\mbox{ constant}.\label{dbc80}
\end{eqnarray}
Going to the second order action and using eq.~(\ref{bdygf}) we recover the original model 
describing
open strings on a K\"ahler manifold with K\"ahler potential $V(z+ \bar z)$ in the presence of a 
space-filling B-brane on which one has a 
holomorphic $U(1)$ bundle determined by the prepotential $W(z+ \bar z)$. The boundary conditions
eq.~(\ref{bdy41}) follow from combining eq.~(\ref{bdygf}) with eq.~(\ref{zbzex}).

We now turn to the dual model which one obtains by integrating the first order action (in the second 
form of eq.~(\ref{fo80})) over 
the gauge fields $y^ \alpha $. Doing so, one finds eq.~(\ref{bulkA}) as the bulk 
equations of motion. It implicitly gives the $y^ \alpha $'s as a function of the twisted chiral 
superfields $w^ \alpha + w^{ \bar \alpha }$. Passing from the first order action eq.~(\ref{fo80}) 
to the second order action, we get the action for the dual model,
\begin{eqnarray}
{\cal S}=\int d^2 \sigma d^2 \theta D'\hat D' \hat V( X, \bar X)+ i\,\int d \tau d^2 \theta\,
\hat W(X, \bar X).\label{aaa80}
\end{eqnarray}
The resulting model is once more K\"ahler with the 
K\"ahler potential given by,
\begin{eqnarray}
\hat V( w+ \bar w)= -V(y(w+ \bar w))+(w_ \alpha +
w_{ \bar \alpha })\,y^ \alpha (w+ \bar w). 
\end{eqnarray}
The K\"ahler metric of the dual model is the inverse of the K\"ahler metric of the original model,
\begin{eqnarray}
\frac{ \partial ^2 \hat V}{ \partial w_ \alpha \partial w_{ \bar \beta }}=
\bigg(\frac{ \partial ^2 V}{ \partial y^ \alpha \partial y^{ \beta }}\bigg)^{-1}
\bigg|_{y=y(w+ \bar w)}.
\end{eqnarray} 
The boundary potential is given by,
\begin{eqnarray}
\hat W( w+ \bar w)&=&W(y(w+ \bar w))-y^ \alpha \, 
\frac{ \partial W(y)}{ \partial y^ \alpha }\Big|_{y=y(w+ \bar w) } \nonumber\\
&=& W(y( w + \bar w))- \frac{ \partial \hat V}{ \partial w_{ \alpha }}
\bigg(\frac{ \partial ^2 \hat V}{ \partial w_ \alpha \partial w_{ \bar \beta }}\bigg)^{-1}
\frac{ \partial  W}{ \partial w_{ \bar \beta  }}\,.
\end{eqnarray}
The model has Dirichlet boundary conditions given by eq.~(\ref{dbc80}) which can be rewritten as,
\begin{eqnarray}
-i\big(w_ \alpha -w_{ \bar \alpha }\big)-
\bigg(\frac{ \partial ^2 \hat V}{ \partial w_ \alpha \partial w_{ \bar \beta }}\bigg)^{-1}\,
\frac{ \partial W(y(w+ \bar w))}{ \partial w_{ \bar \beta } }=\mbox{ constant},\label{dbc83}
\end{eqnarray}
and a set of Neumann boundary 
conditions which either follow from eq.~(\ref{dbc80}) using the constraints eq.~(\ref{n2tcsf})
or which can be obtained by acting with $D'$ and $\hat D$ on eq.~({\ref{bulkA}}) and using 
eq.~(\ref{bdygf}). One verifies that the boundary term in the variation of the action (see 
eq.~(\ref{caut2})) indeed vanishes.

\subsection{From A to B branes}
\subsubsection{Dualizing lagrangian branes}
We start from the D1-brane discussed in section 4, assuming the existence of an isometry. 
The $ \sigma $-model is parametrized by a single 
twisted chiral field $w$ (and its complex conjugate $ \bar w$) with K\"ahler potential 
$V(w+ \bar w)$ and boundary potential $W( w +\bar w)$. So we have $V_w=V_{ \bar w}$ and 
$W_w=W_{ \bar w}$. 
The action is given in eq.~(\ref{aaa0}) and the Dirichlet boundary condition is
\begin{eqnarray}
(V_w+i\,W_w)\, \delta w=(V_{ \bar w}-i\,W_{ \bar w})\, \delta \bar w.\label{ulx1}
\end{eqnarray}
The Neuman boundary condition which follows from this is,
\begin{eqnarray}
(V_w+i\,W_w)\, D' w+(V_{ \bar w}-i\,W_{ \bar w})\, D' \bar w=0.\label{ulx2}
\end{eqnarray}
We introduce a real gauge (unconstrained) superfield $y$ satisfying the boundary condition,
\begin{eqnarray}
\ID 'y=-i\, \frac{W_y(y)}{V_y(y)}\,\ID y,\qquad
\bar \ID 'y=+i\, \frac{W_y(y)}{V_y(y)}\, \bar \ID y.\label{qqbcy}
\end{eqnarray}
The first order action is given by,
\begin{eqnarray}
{\cal S}&=&\int d^2 \sigma \, d^2 \theta \,D'\hat D'\,\Big\{
V(y)-i\,u \,\bar \ID\bar\ID 'y -i \,\bar  u \,\ID\ID'y \Big\}+ \nonumber\\
&&i\int d \tau \, d^2 \theta \bigg\{W(y)+
\bar \ID ' u \,\Big(\bar \ID' y  
-i\, \frac{W_y(y)}{V_y(y)}\, \bar \ID y
\Big)-
\ID ' \bar  u\,\Big( \ID' y
+i\, \frac{W_y(y)}{V_y(y)}\,\ID y
\Big)\bigg\},\label{ga3}
\end{eqnarray}
where the Lagrange multipliers $u $ and $ \bar u=u^\dagger$ are unconstrained
complex $N=(2,2)$ superfields.
Integrating over the Lagrange multipliers yields a bulk term,
\begin{eqnarray}
\bar \ID \bar \ID'\,y =\ID\ID'y =0,
\end{eqnarray}
which is solved in terms of a twisted chiral superfield $w$,
\begin{eqnarray}
y = w+ \bar w.\label{ga4}
\end{eqnarray}
{}From the last term in eq.~(\ref{ga3}) we get a boundary condition as well which is equal to the
one in eq.~(\ref{qqbcy}). Combining the boundary condition with eq.~(\ref{ga4}) and the bulk 
constraints,
\begin{eqnarray}
&&\ID w =+\ID'w ,\quad \bar \ID w=- \bar \ID'w , \nonumber\\
&& \ID \bar w=-  \ID' \bar w,\quad\bar  \ID \bar w =+ \bar \ID' \bar w ,
\end{eqnarray}
which are equivalent to eq.~(\ref{btcf}),
gives the original boundary conditions eqs.~(\ref{ulx1}) and (\ref{ulx2}). 
Going to the second order action
one recovers the original model.

We introduce a potential $Q(y)$ implicitely defined by,
\begin{eqnarray}
W(y)=Q(y)- \frac{V'(y)Q'(y)}{V''(y)}\,,
\end{eqnarray}
where primes denote derivatives with respect to $y$. Using this and partial 
integration\footnote{The calculations are facilitated by using $\int d^2 \sigma \, 
d^2 \theta
\,D'\hat D'=-(1/4)\int d^2 \sigma \, \ID \bar \ID \ID ' \bar \ID '$ and $\int d \tau \,d^2 \theta
=-(i/2)\int d \tau \, \ID \bar \ID $. Once again we refer to appendix A for conventions.}, 
we can rewrite eq.~(\ref{ga3}) as,
\begin{eqnarray}
{\cal S}&=&\int d^2 \sigma \, d^2 \theta \,D'\hat D'\,\Big\{
V(y)-y \left( z+ \bar z\right) \Big\} \nonumber\\
&&+i\int d \tau \,d^2 \theta\, \bigg\{
Q(y)- \frac{V'(y)Q'(y)}{V''(y)}+ \frac{Q'(y)}{V''(y)}\,\big(z+ \bar z\big)
\bigg\},\label{ga97}
\end{eqnarray}
where we introduced the chiral superfield $z$,
\begin{eqnarray}
z\equiv i\,\bar \ID \bar \ID'\,u,\qquad \bar z\equiv i\, \ID  \ID'\, \bar u,\label{ga7}
\end{eqnarray}
which by construction satisfy the constraints,
\begin{eqnarray}
\bar \ID z = \bar \ID 'z =\ID \bar z=\ID' \bar z=0.\label{ga8}
\end{eqnarray}
Integrating over the unconstrained superfield $y$ gives the bulk equation of
motion,
\begin{eqnarray}
z+ \bar z= \frac{ \partial\, V(y)}{ \partial \,y}\,,\label{ewoi}
\label{ga9}
\end{eqnarray}
which upon inversion gives $y $ as a function of $ z+ \bar z  $: $y=y( z+ \bar z)$.
The boundary term arising from varying $y$,
\begin{eqnarray}
\delta {\cal S}\Big|_{boundary}=-i\int d \tau\,d^2 \theta \, \delta y\,
\left(
\frac{Q'(y)}{V''(y)}
\right)' \,\big(
V'(y)-(z+ \bar z)
\big),
\end{eqnarray}
vanishes by virtue of eq.~(\ref{ewoi}).
Using eqs.~(\ref{ga9}) and (\ref{ga8}), we get from eq.~(\ref{qqbcy}) the Neumann
boundary conditions,
\begin{eqnarray}
\ID'z&=&-i\, \frac{W'\big(y( z + \bar z)\big)}{V'\big(y(z+ \bar z)\big)}\,
\ID z, \nonumber\\
\bar \ID' \bar z&=&+i\, \frac{W'\big(y( z + \bar z)\big)}
{V'\big(y(z+ \bar z)\big)}\, \bar \ID \bar z.\label{ga88}
\end{eqnarray}

We now go to the second order action. In order to make this as explicit as possible, we introduce a
potential $P(y)$ defined by
\begin{eqnarray}
V(y)=-\int dy\,P(y).
\end{eqnarray}
With this eq.~(\ref{ga9}) can be rewritten as,
\begin{eqnarray}
z+ \bar z= P(y),
\end{eqnarray}
or,
\begin{eqnarray}
y=P^{-1}(z + \bar z).
\end{eqnarray}
Using this, the second order action follows from eq.~(\ref{ga97}):
\begin{eqnarray}
{\cal S}&=&-\int d^2 \sigma \, d^2 \theta \,D'\hat D'\,\int d\big(z+ \bar z\big)\,
P^{-1}(z+ \bar z)+i\,\int d \tau \,d^2 \theta \,Q\big(P^{-1}(z + \bar z)\big),
\end{eqnarray}
from which we read the K\"ahler potential $\hat V( z+ \bar z)$ and the $U(1)$ potential
$\hat W( z + \bar z)$:
\begin{eqnarray}
\hat V( z+ \bar z)=\int d(z + \bar z)\,P^{-1}(z + \bar z),\qquad
\hat W( z+ \bar z)=Q\big(P^{-1}(z+ \bar z)\big).
\end{eqnarray}
In terms of the dual variables we can rewrite the boundary conditions eq.~(\ref{ga88}) as,
\begin{eqnarray}
D'z=+i\, \frac{\hat W_{z \bar z}}{ \hat V_{z \bar z}}\,Dz,\qquad
D' \bar z=-i\, \frac{\hat W_{z \bar z}}{ \hat V_{z \bar z}}\, \bar Dz,
\end{eqnarray}
which are recognized as the standard Neumann boundary conditions in the presence of magnetic 
background field.

Concluding we find that the dual theory describes open strings in the presence of a space filling
D$2$ B-brane on a K\"ahler manifold whose potential is given by 
$ \hat V=\int d(z + \bar z)\,P^{-1}(z+ \bar z)$. In addition,
a $U(1)$ bundle with potential $ \hat W=Q\big(P^{-1}(z + \bar z)\big)$ is present as well.

\subsubsection{Dualizing coisotropic branes}
Now let us look at the case of coisotropic A branes. The example discussed at the end of section 4 
is characterized by a K\"ahler potential $V(z- \bar z, w+ \bar w)$ which satisfies the 
Monge-Amp\`ere equation eq.~(\ref{MAeq}). The potential has an obvious isometry,
\begin{eqnarray}
\delta z= -\varepsilon _1,\qquad \delta w= -i \, \varepsilon _2,\label{iso9}
\end{eqnarray}
with $ \varepsilon _1,\ \varepsilon _2\in\IR $ 
and constant.  However the boundary potential $W$ given in 
eq.~(\ref{exWex}) can be rewritten as,
\begin{eqnarray}
W= \frac{i}{2}\Big(\big(z+ \bar z\big)\,V_z(z- \bar z, w+ \bar w)+
\big(w- \bar w\big)\,V_w(z- \bar z, w+ \bar w)
\Big),\label{bpot77}
\end{eqnarray}
and does not exhibit the above mentioned isometry. Remarkably one finds -- using the fact
that the K\"ahler potential satisfies the Monge-Amp\`ere equation eq.~(\ref{MAeq}) -- that
$\hat D\,W$ transforms in a total $D$ 
derivative, making the boundary term in the action invariant as well. Let us make this very
explicit by making a change of coordinates:
\begin{eqnarray}
&&z_1\equiv z+ \bar z-i\,V_w ,\qquad \bar z_1\equiv z+ \bar z+iV_w, \nonumber\\
&&z_2\equiv -i\big(w- \bar w\big)+V_z,\qquad \bar z_2\equiv 
-i\big(w- \bar w\big)-V_z.
\end{eqnarray}
One verifies that both $z_1$ and $z_2$ are chiral boundary fields, {\em i.e.},
\begin{eqnarray}
\hat Dz_a=+iDz_a,\qquad \hat D \bar z_a=-iD \bar z_a\label{cf55} 
\end{eqnarray}
for $a\in\{1,2\}$. The boundary potential $W$ given in eq.~(\ref{bpot77}) can be rewritten as,
\begin{eqnarray}
W= \frac{i}{8}\Big(
\big(z_1+ \bar z_1\big)\big(z_2- \bar z_2\big)-
\big(z_1- \bar z_1\big)\big(z_2+ \bar z_2\big)
\Big).\label{bpot78}
\end{eqnarray}
The isometry eq.~(\ref{iso9}) becomes in these coordinates
\begin{eqnarray}
\delta z_1= \delta \bar z_1= -2\,\varepsilon _1,\qquad 
\delta z_2= \delta \bar z_2= -2 \, \varepsilon _2.\label{iso10}
\end{eqnarray}
Under these transformations, the potential transforms as
\begin{eqnarray}
\delta W= - \frac{i}{2}\Big(
\big( \varepsilon _1z_2- \varepsilon _2z_1\big)-\big(
\varepsilon _1 \bar z_2- \varepsilon _2 \bar z_1
\big)
\Big),
\end{eqnarray}
which -- by virtue of the constraints eq.~(\ref{cf55}) -- gives 
$ \delta \int d \tau d^2 \theta \,W=0$. The present situation is similar to the one studied in
\cite{Hull:1985pq}. In order to gauge the isometries, one needs first to modify the potential
$W$ such that it becomes invariant under the isometries. This is achieved by modifying $W$ to $W'$,
\begin{eqnarray}
W'=W+ \frac{i}{2}\Big(q\, z_1z_2-q\, \bar z_1 \bar z_2+ \xi - \bar \xi \Big),
\end{eqnarray}
where $q\in \IR$ and $ \xi $ is a new (auxiliary) boundary {\em chiral} field which transforms under
the isometry as,
\begin{eqnarray}
\delta \xi= (1+2q) \,\varepsilon _1z_2-(1-2q)\, \varepsilon _2 z_1.\label{iso11}
\end{eqnarray}
With this one gets that $ \delta W'=0$.
Because the difference between $W$ and $W'$ is the sum of a holomorphic and an anti-holomorphic
function of the boundary chiral fields we have that 
$\int d \tau d^ 2 \theta \,W'=\int d \tau d^ 2 \theta \,W$, so the physical content of the model 
remains unchanged. However -- as was shown in \cite{Hull:1985pq} -- when trying to gauge more than 
one isometry simultanously one can encounter an obstruction (which was given a Lie algebra
cohomology interpretation in \cite{de Wit:1987ph}) which 
renders gauging of the full isometry algebra 
impossible. In the present situation this obstruction is indeed present -- as one can check using
the equations 
developed in \cite{Hull:1985pq} -- implying that we can only gauge a linear combination 
of the isometries given in eqs.~(\ref{iso10}) and (\ref{iso11}). 

{}For concreteness, we will gauge the $ \varepsilon _2$ isometry. Our analysis is considerably 
simplified if we rewrite the boundary term in the action as,
\begin{eqnarray}
{\cal S}\Big|_{boundary}&=&
i\int d \tau d^2 \theta \,W=
i\int d \tau d^2 \theta \, \frac{i}{4}\big(z_1+ \bar z_1\big)\big(z_2- \bar z_2\big) \nonumber\\
&=&i\int d \tau d^2 \theta \, i\,\big( z+ \bar z\big)\,V_z.
\end{eqnarray} 
The gauging procedure is now simple. We introduce an unconstrained gauge field $y$ satisfying the
boundary conditions,
\begin{eqnarray}
&&\ID ' y=+i\,\ID V_z(z- \bar z,y),\qquad
\bar \ID ' y=+i\, \bar \ID V_z(z- \bar z,y), \label{ioy1}\\
&&\ID'\big(z - \bar z\big)=-i\,\ID V_y(z- \bar z,y),\qquad
\bar \ID'\big(z - \bar z\big)=-i\, \bar \ID V_y(z- \bar z,y).\label{ioy2}
\end{eqnarray}
The first order action is given by,
\begin{eqnarray}
{\cal S}&=&\int d^2 \sigma \, d^2 \theta \,D'\hat D'\,\Big\{
V(z- \bar z,y)-i\,u \,\bar \ID\bar\ID 'y -i \,\bar  u \,\ID\ID'y \Big\} \nonumber\\
&&\qquad+i\int d \tau \, d^2 \theta \Big\{i\big(z + \bar z\big)V_z(z- \bar z,y)+
\bar \ID ' u \,\Big(\bar \ID' y  
-i\, \bar \ID V_z(z- \bar z,y)
\Big) \nonumber\\
&&\qquad-\ID ' \bar  u\,\Big( \ID' y
-i\, \ID V_z( z- \bar z,y)
\Big)\Big\},\label{ggaa3}
\end{eqnarray}
where $u$ and $ \bar u\equiv u^\dagger$ are unconstrained complex $N=(2,2)$ superfields.
Integrating over $u$ and $ \bar u$ gives the equation of motion,
\begin{eqnarray}
\ID\ID'y= \bar \ID \bar \ID'y=0,
\end{eqnarray}
which is solved by putting, 
\begin{eqnarray}
y= w+ \bar w,
\end{eqnarray}
with $w$ a twisted chiral superfield. Varying $y$ yields a boundary term as well which vanishes if 
we impose eq.~(\ref{ioy1}). So the action eq.~(\ref{ggaa3}) together with the boundary condition 
eq.~(\ref{ioy2}) reproduces upon integrating over $u$ and $ \bar u$ the original theory. 

Partially integrating, we rewrite eq.~(\ref{ggaa3}) as,
\begin{eqnarray}
{\cal S}&=&\int d^2 \sigma \, d^2 \theta \,D'\hat D'\,\Big\{
V(z- \bar z,y)-y\,\big(r+ \bar r\big)\Big\} \nonumber\\
&&\qquad +i\,\int d \tau \, d^2 \theta \Big\{i\big(z + \bar z\big)
-\big(r- \bar r\big)\Big\}\, V_z(z- \bar z,y),\label{ggaa5}
\end{eqnarray}
where we introduced the chiral field $r$ (and $ \bar r$),
\begin{eqnarray}
r\equiv i\, \bar \ID \bar \ID'u,\qquad \bar r\equiv i\,\ID\ID' \bar u.
\end{eqnarray}
Integrating over $y$ yields the equation of motion,
\begin{eqnarray}
V_y( z- \bar z,y)= r + \bar r.\label{ggaa7}
\end{eqnarray}
In terms of the prepotential $h$ introduced in eq.~(\ref{exLT}), we can write a second order
expression for the integrand of 
the bulkterm in eq.~(\ref{ggaa5}) as,
\begin{eqnarray}
\Big(V(z- \bar z,y)-y\,\big(r+ \bar r\big)\Big)\Big|_{y=y(z- \bar z, r+ \bar r)}=
h(z - \bar z- r - \bar r)+ \bar h( \bar z-z-r - \bar r)\,.\label{ggaa11}
\end{eqnarray}
Furthermore, requiring that the boundary term in the variation of $y$
vanishes gives the Dirichlet boundary condition,
\begin{eqnarray}
\mbox{Im}\,r-\mbox{Re}\,z=0.\label{lq1}
\end{eqnarray}
Combining eqs.~(\ref{ioy2}) and (\ref{ggaa7}) yields a Dirichlet,
\begin{eqnarray}
\mbox{Im}\,r-\mbox{Re}\,z=\mbox{constant},\label{lq2}
\end{eqnarray}
and a Neumann,
\begin{eqnarray}
-i\,\big(D'z-D' \bar z\big)=-D\big(r + \bar r\big),
\end{eqnarray}
boundary condition. Note that eqs.~(\ref{lq1}) and (\ref{lq2}) are mutually compatible
if we choose the constant in eq.~(\ref{lq2}) to be zero. Finally,
the combination of eq.~(\ref{ioy1}) and (\ref{ggaa7}) yields two more Neumann boundary
conditions,
\begin{eqnarray}
&&-i\,\big(D'r-D' \bar r\big)+D'z+D' \bar z=0, \nonumber\\
&&D'r+D' \bar r=-i\,D\big(z- \bar z\big).
\end{eqnarray}
So this implies that the open strings are propagating in a background which contains a D3-brane 
whose location is fixed by eq.~(\ref{lq1}). The bulk geometry is bi-hermitean and parametrized by
a chiral, $r$, and a twisted chiral field, $z$, with the generalized K\"ahler potential given by
$h(z - \bar z- r - \bar r)+ \bar h( \bar z-z-r - \bar r)$. The non-vanishing components of the 
metric and the Kalb-Ramond form can be obtained from eq.~(\ref{KRform}) and are given by, 
\begin{eqnarray}
&&g_{r \bar r}=g_{z \bar z}=+h''(z- \bar z-r- \bar r)+
\bar h''( \bar z-z-r- \bar r), \nonumber\\
&&b_{r \bar z}=g_{z \bar r}=-h''(z- \bar z-r- \bar r)+
\bar h''( \bar z-z-r- \bar r).
\end{eqnarray}
Models whose bulk geometry 
is generalized K\"ahler will be studied in detail 
in \cite{wip}. Nonetheless, the previous example 
clearly shows that coisotropic branes do appear as duals to other brane configurations.
\section{Conclusions and discussion}

In this paper we presented a completely local $N=2$ superspace
formulation of two-dimensional nonlinear $ \sigma $-models for
target spaces parameterized exclusively by chiral or twisted chiral
fields (meaning that the bulk geometry is K\"ahler). 
This was possible because, contrary to previous attempts,
only the supersymmetries which are preserved by the boundary
conditions were required to remain manifest at all times.
Starting from this formalism, a general $N=2$
superspace description of both A and B branes on K\"ahler manifolds
was given. Interchanging type A boundary conditions for type B and vice versa
turns out to be equivalent to exchanging chiral for twisted chiral
superfields and vice versa allowing us without loss of generality to limit
ourselves to type B boundary conditions. In this setting A-branes (B-branes) are described solely
in terms of twisted-chiral (chiral) superfields.

The $N=2$ superspace description of type A branes turned out to be subtle.
It gives rise to a ``non-standard" boundary
coupling which was shown to reduce to the standard one when
proper use is made of the nontrivial boundary conditions. An open question
-- for the case of A-branes -- is to find a better characterization or geometric
interpretation of the boundary potential $W$. Perhaps a reformulation of the problem in terms of 
generalized complex geometry might shed some light.

The duality transformations relating A and B models
were investigated as well. The main difficulty here is the identification of the right
boundary terms in the first order action which see to it that boundary conditions correctly
carry over during the duality transformation. When isometries are present, it is reasonably 
straightforward to dualize lagrangian A-branes to space filling B-branes and vice-versa. 
Dualizing a coisotropic A-brane turns out to be subtle. The example of a space-filling D4 
coisotropic brane was shown to have two isometries. However only a linear combination of 
those two can be gauged. As a consequence we can dualize the model to a D3-brane where the bulk
is now not K\"ahler anymore, but exhibits a bihermitean geometry. 

It is clear that in general not sufficient isometries will be present to convert an A brane on a 
K\"ahler manifold to a B-brane on a K\"ahler manifold or vice-versa. When only part of the chiral or 
twisted chiral superfields can be dualized, the dual model will exhibit a bihermitian -- or
equivalently, a generalized K\"ahler -- geometry. The study of these duality transformations will be 
reported on in \cite{wip}.

Finally, $N=2$ superspace provides a powerful framework for investigating the quantum properties
of these non-linear $ \sigma $-models (as was demonstrated in {\em e.g.}~\cite{Nevens:2006ht}). 
Requiring the
$ \beta $-functions to vanish gives rise to further conditions on the background geometry. E.g.~at 
one loop one gets that the holomorphic bundle for a type B brane needs to satisfy a deformed
stability condition as well. 
In this context it would be most interesting to calculate the one loop $ 
\beta $-function for a coisotropic brane and make contact with the stability conditions obtained in
\cite{Kapustin:2003se}. 

\acknowledgments

\bigskip

We thank Peter Bouwknegt, Frederik Denef, Chris Hull, Paul Koerber, Lucca Martucci,
Martin Ro\v cek and especially Ulf Lindstr\"om for useful discussions and suggestions.
All authors are supported in part by the Belgian Federal Science Policy Office
through the Interuniversity Attraction Pole P6/11, in part by the European
Commission FP6 RTN programme MRTN-CT-2004-005104 and in part by the
``FWO-Vlaanderen'' through project G.0428.06.

\appendix

\section{Conventions, notations and identities}\label{app conv}
We denote the worldsheet coordinates by $ \tau \in\IR$ and $ \sigma \in \IR$, $ \sigma \geq 0$.
Sometimes we use worldsheet light-cone coordinates,
\begin{eqnarray}
\sigma ^\pp= \tau + \sigma ,\qquad \sigma ^== \tau - \sigma .\label{App1}
\end{eqnarray}
The $N=(1,1)$ (real) fermionic coordinates are denoted by $ \theta ^+$ and $ \theta ^-$ and the
corresponding derivatives satisfy,
\begin{eqnarray}
D_+^2= - \frac{i}{2}\, \partial _\pp \,,\qquad D_-^2=- \frac{i}{2}\, \partial _= \,,
\qquad \{D_+,D_-\}=0.\label{App2}
\end{eqnarray}
Passing from $N=(1,1)$ to $ N=(2,2)$ superspace requires
the introduction of two more real fermionic coordinates $ \hat \theta ^+$ and $ \hat \theta ^-$
where the corresponding fermionic derivatives satisfy,
\begin{eqnarray}
\hat D_+^2= - \frac{i}{2} \,\partial _\pp \,,\qquad \hat D_-^2=- \frac{i}{2} \,\partial _= \,,
\end{eqnarray}
and again all other -- except for (\ref{App2}) -- (anti-)commutators do vanish. Quite often a complex basis is used,
\begin{eqnarray}
\ID_\pm\equiv \hat D_\pm+i\, D_\pm,\qquad
\bar \ID_\pm\equiv\hat D_\pm-i\,D_\pm,
\end{eqnarray}
which satisfy,
\begin{eqnarray}
\{\ID_+,\bar \ID_+\}= -2i\, \partial _\pp\,,\qquad
\{\ID_-,\bar \ID_-\}= -2i\, \partial _=,
\end{eqnarray}
and all other anti-commutators do vanish.

When dealing with boundaries in $N=(2,2)$ superspace, we introduce
various derivatives as linear combinations of the previous ones. We
summarize their definitions together with the non-vanishing
anti-commutation relations. We have,
\begin{eqnarray}
&&D\equiv D_++D_-,\qquad \hat D\equiv \hat D_++ \hat D_-, \nonumber\\
&& D'\equiv D_+-D_-,\qquad \hat D'\equiv \hat D_+- \hat D_-,
\end{eqnarray}
with,
\begin{eqnarray}
&&D^2=\hat D^2=D'{}^2=\hat D'{}^2=- \frac{i}{2} \partial _ \tau , \nonumber\\
&&\{D,D'\}=\{\hat D, \hat D'\}=-i \partial _ \sigma .
\end{eqnarray}
In addition we also use,
\begin{eqnarray}
&&\ID\equiv \ID_++\ID_-=\hat D+i\,D,\qquad \ID'\equiv \ID_+-\ID_-=\hat D'+i\,D', \nonumber\\
&& \bar \ID\equiv \bar \ID_++ \bar \ID_-=\hat D-i\,D,\qquad \bar \ID'\equiv \bar \ID_+- \bar
\ID_-=\hat D'-i\,D'.
\end{eqnarray}
They satisfy,
\begin{eqnarray}
&&\{\ID, \bar \ID\}=\{\ID', \bar \ID'\}= -2i\, \partial _ \tau ,\, \nonumber\\
&&\{\ID, \bar \ID'\}=\{\ID', \bar \ID\}=-2i\, \partial _ \sigma  \,.
\end{eqnarray}

\section{Submanifolds of symplectic manifolds}\label{app subm}

A symplectic manifold ${\cal M}$ is a manifold endowed with a
non-degenerate closed two-form $\omega$. There are several natural
ways to define specific submanifolds of these. We will do this by
first defining the symplectic complement of a subspace of a
symplectic vector space.

So let $V$ be a symplectic vector space of dimension $d=2n$. This
means that it is equipped with a non-degenerate, skew-symmetric,
bilinear form $\omega$, called the symplectic form. The symplectic
complement of a subspace $W$ is defined as,
\begin{eqnarray}
W^\bot = \{v\in V \vert \omega(v,w) = 0, \forall w \in W\}.
\end{eqnarray}
This satisfies $(W^\bot)^\bot = W$ and $\dim W + \dim W^\bot = \dim
V$. However, contrary to the orthogonal complement (defined with a
metric), generically $W \cap W^\bot \neq \emptyset$.

We are interested in the three following cases,
 \begin{description}
 \item[Isotropic: ] When $W \subseteq W^\bot$, $W$ is called isotropic.
 This is true if and only if $\omega$ restricts to zero on $W$. Every
 one-dimensional subspace is isotropic.
 \item[Coisotropic: ] When $W^\bot \subseteq W$, $W$ is called
 coisotropic. In other words, $W$ is coisotropic if and only if
 $W^\bot$ is isotropic. Equivalently, $W$ is coisotropic if and only
 if $\omega$ descends to a non-degenerate form on the quotient space
 $W/W^\bot$. A codimension one subspace is always coisotropic.
 \item[Lagrangian: ] When $W = W^\bot$, $W$ is called Lagrangian, so
 that a Lagrangian subspace is both isotropic and coisotropic.
 \end{description}
These definitions immediately imply that, because of the
non-degeneracy of $\omega$, a Lagrangian subspace is
$n$-dimensional, where $n = d /2$. The number of dimensions of an
isotropic (a coisotropic) subspace in necessarily smaller (bigger)
than $n$.

Given a symplectic manifold ${\cal M}$, a submanifold ${\cal N}$ is
called isotropic, coisotropic or Lagrangian if the tangent space
$T_{\cal N}$ is an isotropic, coisotropic or Lagrangian subspace of
$T_{\cal M}$, that is, if $T_{\cal N} \subseteq T_{\cal N}^\bot$,
$T_{\cal N}^\bot \subseteq T_{\cal N}$ or $T_{\cal N} = T_{\cal
N}^\bot$, respectively.

\end{document}